\documentclass[%
 reprint,
 amsmath,amssymb,
 aps,
]{revtex4-1}

\usepackage{graphicx}% Include figure files
\usepackage{dcolumn}% Align table columns on decimal point
\usepackage{bm}% bold math
\usepackage{xspace}
\usepackage{multirow}
\usepackage{booktabs}
\usepackage{xcolor}
\usepackage{xfrac}
\usepackage[caption=false]{subfig}
\graphicspath{{./plots/}}

\usepackage{hyperref}
\hypersetup{
    colorlinks=true,
    linkcolor=blue,
    filecolor=blue,      
    urlcolor=blue,
    pdftitle={Overleaf Example},
    pdfpagemode=FullScreen,
    }

\usepackage{units}
\usepackage{comment}

\def\Sem{$S_{\rm em}$\xspace}
\def\Smu{$S_{\rm \mu}$\xspace}
\def\d{{\rm d}}

\def\nc{M_{\nu_e}(\mathrm{NC})}%
\def\cc{M_{\nu_e}(\mathrm{CC})}%
\def\snc{\sigma_{\mathrm{NC}}}%
\def\scc{\sigma_{\mathrm{CC}}}%

\newcommand{\x}[1]{%
  {}$% get out of math
  \kern-2\mathsurround % in case it's non zero
  $% reenter math
  \binoppenalty10000 \relpenalty10000 #1% typeset the subformula
  {}$% get out of math
  \kern-2\mathsurround % in case it's non zero
  $% reenter math for the rest of the formula
}

\begin{document}

%\preprint{APS/123-QED}

\title{Evaluation of the potential of a gamma-ray observatory to detect astrophysical neutrinos through inclined showers}% Force line breaks with \\

\author{Jaime Alvarez-Mu\~niz}
\address{Instituto Galego de F\'\i{}sica de Altas Enerx\'\i{}as (IGFAE), Universidade de Santiago de Compostela, 15782 Santiago de Compostela, Spain}

\author{Ruben Concei\c{c}\~{a}o}
\email{ruben@lip.pt}
\address{Laborat\'{o}rio de Instrumenta\c{c}\~{a}o e F\'{i}sica Experimental de Part\'{i}culas (LIP) - Lisbon, Av.\ Prof.\ Gama Pinto 2, 1649-003 Lisbon, Portugal}
\address{Instituto Superior T\'ecnico (IST), Universidade de Lisboa, Av.\ Rovisco Pais 1, 1049-001 Lisbon, Portugal}

\author{Pedro J. Costa}
\address{Laborat\'{o}rio de Instrumenta\c{c}\~{a}o e F\'{i}sica Experimental de Part\'{i}culas (LIP) - Lisbon, Av.\ Prof.\ Gama Pinto 2, 1649-003 Lisbon, Portugal}
\address{Instituto Superior T\'ecnico (IST), Universidade de Lisboa, Av.\ Rovisco Pais 1, 1049-001 Lisbon, Portugal}

\author{M\'ario Pimenta}
\address{Laborat\'{o}rio de Instrumenta\c{c}\~{a}o e F\'{i}sica Experimental de Part\'{i}culas (LIP) - Lisbon, Av.\ Prof.\ Gama Pinto 2, 1649-003 Lisbon, Portugal}
\address{Instituto Superior T\'ecnico (IST), Universidade de Lisboa, Av.\ Rovisco Pais 1, 1049-001 Lisbon, Portugal}

\author{Bernardo Tom\'e}
\address{Laborat\'{o}rio de Instrumenta\c{c}\~{a}o e F\'{i}sica Experimental de Part\'{i}culas (LIP) - Lisbon, Av.\ Prof.\ Gama Pinto 2, 1649-003 Lisbon, Portugal}
\address{Instituto Superior T\'ecnico (IST), Universidade de Lisboa, Av.\ Rovisco Pais 1, 1049-001 Lisbon, Portugal}

\date{\today}% It is always \today, today,

\begin{abstract}
  We assess the capabilities of a ground-based gamma-ray observatory to detect astrophysical neutrinos with energies in the $100\,{\rm TeV}$ to $100\,{\rm PeV}$ range. The identification of these events would be done through the measurement of very inclined extensive air showers induced by downward-going and upward-going neutrinos. The discrimination of neutrino-induced showers in the overwhelming cosmic-ray background is achieved by analysing the balance of the total electromagnetic and muonic signals of the shower at the ground. We demonstrate that a ${\rm km^2}$-scale wide field-of-view ground-based gamma-ray observatory could detect a couple of Very-High to Ultra-High energy (VHE-UHE) neutrino events per year with a reasonable pointing accuracy, making it an interesting facility for multi-messenger studies with both photons and neutrinos.
\end{abstract}

\pacs{Valid PACS appear here}% PACS, the Physics and Astronomy
\maketitle

%\tableofcontents

%%%%%%%%%%%%%%%%%%%%%%%%%%%%%%%%%%%%%%%%%%%%%%%%%%%%%%%%%%%%%%%%%%%%%%%%%%
\section{Introduction}
\label{sec:intro}

The multi-messenger approach to astroparticle physics has the potential to address fundamental problems, such as those related to physics in extreme phenomena, the origin of ultra-high-energy cosmic rays, the nature of dark matter, the possibility of Lorentz invariance violation, and even the existence of undiscovered particles. 

Numerous experiments resort to extensive air shower (EAS) arrays to study very-high-energy gamma-rays, such as HAWC~\cite{HAWC}, LHAASO~\cite{LHAASOLayout}, and the Southern Wide-field Gamma-ray Observatory (SWGO)~\cite{SWGOFuture}, currently in its planning stage. The recent observation of gamma-rays with energies above $1\,$PeV by LHAASO~\cite{LHAASOPeV} puts pressure on the construction of a facility surveying the Southern hemisphere sky. This experiment should have an effective area of the order of ${\rm km^2}$ and an excellent gamma/hadron discrimination capabilities to cope with the low fluxes reported by LHAASO.

On the other hand, experiments such as IceCube have been successfully operating over the years, demonstrating the presence of a very-high-energy neutrino flux of astrophysical origin. This flux has been seen to extend up to a few PeV with no sign of a cutoff~\cite{IceCubeFlux}.

The simultaneous measurement of gamma-rays and neutrinos coming from the same astrophysical source, known as multi-messenger measurements, is highly aspired, and it has in the last years been reshaping the experimental panorama with the addition of new, more ambitious upgrades and new experiments (see, for instance, \cite{IceCube-Gen2,CTAFundPhys,FermiGW}).

In this work, we have used shower simulations to determine whether ground-based gamma-ray EAS arrays can be used to detect neutrinos and estimate their expected sensitivity. Our study is restricted to neutrinos with energies ranging from $100\,{\rm TeV}$ to $100\,{\rm PeV}$. Signal events correspond to inclined EAS (zenith angle $\theta>60^\circ$) induced by downward and upward-going neutrinos. The main background source for this measurement is very inclined EAS resulting from the interaction of cosmic rays with the atmosphere.

The article is organized as follows: In section~\ref{sec:strategy}, the experimental strategy employed to distinguish showers induced by neutrinos from the cosmic ray background is presented. Next, in Section~\ref{sec:simulation}, the simulation framework and the sets of simulated showers are given. In Section~\ref{sec:discrimination}, the discrimination methodology is presented. In Section~\ref{sec:method}, we discuss the method to estimate the sensitivity of a ground array observatory to astrophysical neutrinos, focusing on electron neutrinos $\nu_e$. Our results on the sensitivity obtained for downward-going and upward-going neutrino-induced events are given in Sections~\ref{sec:resultsdown} and~\ref{sec:resultsup}, respectively. In Section~\ref{sec:resultsdown}, the impact of the density of detector units in the array (fill factor), of experimental reconstruction resolution, and of simulations statistics are studied. Finally, in Section \ref{sec:resultsall}, an estimate of the sensitivity considering all neutrino flavours is presented. We end the article in Section~\ref{sec:conclusions} with some final remarks and conclusions.

%%%%%%%%%%%%%%%%%%%%%%%%%%%%%%%%%%%%%%%%%%%%%%%%%%%%%%%%%%%%%%%%%%%%%%%%%%
\section{Experimental strategy}
\label{sec:strategy}

In this work, we investigate the sensitivity of a ground-based wide field-of-view gamma-ray observatory, such as the LHAASO experiment~\cite{LHAASOLayout} or the future SWGO~\cite{SWGOFuture}, for the detection of astrophysical neutrinos in the energy range of hundreds of TeV up to hundreds of PeV. These experiments cover large effective areas of $\sim 1\,{\rm km^2}$ with a relatively high fill factor\footnote{In this context, the fill factor is the total detector sensitive area over the shower sampling area (size of the array).} ($\sim 4\%$ for LHAASO) to boost the detection of the very-low photon fluxes at $> \mathrm{PeV}$ energies. 

The main source of background for these observatories is the overwhelming cosmic-ray flux that supersedes the gamma-ray flux by a factor $\sim 10^4$ above $100\,$TeV energy. To mitigate this background, experimental data is often analysed to extract the muon content of the shower, which is higher for hadron-induced showers. However, the distinction between vertical (zenith angle $\theta\lesssim60^\circ$) neutrino-induced and cosmic-ray-induced showers is complicated, as the events exhibit similar signatures. The discrimination is enhanced for inclined showers ($\theta\gtrsim60^\circ$) due to the larger depth of atmosphere between the point of first interaction and the ground~\cite{PierreAuger:2011cpc}. As the proton-air interaction cross-section is seven orders of magnitude larger than the neutrino-air one, protons typically interact in the upper layers of the atmosphere, and a proton-induced inclined shower has to cross a large amount of matter before reaching the ground level. As a consequence, most of the electromagnetic component gets absorbed, and only muons can reach the ground. As a result, ground-based array detectors sample what is commonly called an \emph{old} shower.
Neutrinos, on the other hand, can interact much closer to the detector stations, and both the electromagnetic and muonic components will be detected, what is commonly called a \emph{young} shower. Thus, the balance between the amount of measured signal due to muons and electromagnetic particles can be used to discriminate neutrino from cosmic-ray induced showers.
This strategy has also been used by the surface detector array of the Pierre Auger Observatory to place limits on the neutrino flux at EeV energies~\cite{PAOUHENus,InclinedNusPAO}.

Hence, the neutrino signatures that we investigate in this work are those of very inclined showers ($\theta$ in the range $60^\circ$ to $88^\circ$) initiated close to the ground. Neutrinos with energies in the $100~{\rm TeV}-100~{\rm PeV}$ range are taken as signal, while the background is mainly attributed to very inclined EAS induced by cosmic rays. We initially focus on studying the detection of electron neutrinos $\nu_e$ only. When these particles interact with the atmosphere, they can generate both a hadronic and an electromagnetic shower, maximizing the detection probability. Upon reaching the ground, the inclined cascade may have undergone a substantial development producing a large footprint and facilitating its detection with a surface detector array. 

The key observables to discriminate between neutrino and proton-induced showers are the total amount of signal produced by electromagnetic particles (\Sem) and by muons (\Smu). The existing and planned gamma-ray experiments should be able to access both quantities. The electromagnetic signal is essential to estimate the primary energy, while \Smu is typically used to discriminate gamma from proton-induced showers. In this work, we assume that both quantities are readily available instead of performing a dedicated experiment-dependent reconstruction (see, for instance, the LHAASO experiment~\cite{LHAASOLayout} to see how these quantities can be accessed). Afterwards, in Section~\ref{sec:reconstruction}, the impact of a possible reconstruction uncertainty on the sensitivity to VHE neutrinos is discussed. This study allows for the extraction of the experimental resolution needed to allow the detection of neutrino events.

%%%%%%%%%%%%%%%%%%%%%%%%%%%%%%%%%%%%%%%%%%%%%%%%%%%%%%%%%%%%%%%%%%%%%%%%%%
\section{Simulation Framework and Data Analysis}
\label{sec:simulation}

\begin{figure}[ht]
    \centering
	\includegraphics[width=1.\linewidth]{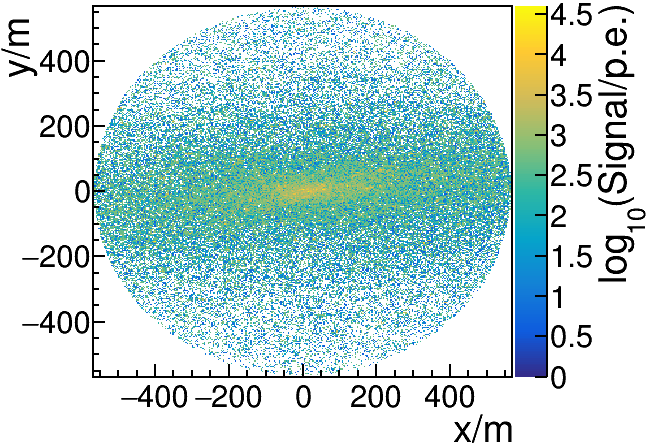}
	\caption{Average footprint of the signal generated by 1000 proton-induced showers of energy $E_p=100\,{\rm TeV}$, zenith angle $\theta=75^\circ$, and azimuthal angle $\phi = 0^\circ$ on a water-Cherenkov detector (WCD) array. The array spans an area of $1\,{\rm km^2}$ with an $80\%$ fill factor. Each WCD station covers an area of $12.6\,{\rm m^2}$. The $x=0$ and $y=0$ corresponds to the projection to the ground of the initial cosmic ray direction.}
	\label{fig:DetectArray}
\end{figure}

We have simulated the development of air showers with dedicated Monte Carlo codes, and assumed a flat EAS array composed of cylindrical water-Cherenkov detector (WCD) units with area $\sim 12\,{\rm m^2}$, spanning over an area of $1\,{\rm km^2}$. The response of the station unit is modeled using a parameterisation of the average signal as a function of the energy of the particle crossing the detector. An example of the average air-shower footprint at the ground is displayed in Fig.~\ref{fig:DetectArray}.

CORSIKA (COsmic Ray Simulations for KAscade - version 7.7410)~\cite{CORSIKA} was used to generate downward-going extensive air showers initiated by protons and neutrinos.
Neutrino-induced air showers were simulated at fixed interaction points from the ground level up to $12\,000\,{\rm m}$ in vertical height, while for proton-induced showers, the starting points were sampled taking into account the proton-air cross-section. Showers generated by upward-going neutrinos interacting within the Earth's crust and developing in the ground, were simulated using the AIRES framework, version 2.8.4a~\cite{AIRES}. Simulations were performed at fixed values of energy and zenith angle, while the azimuth angle ($\phi$) was sampled from a $2\pi$ uniform distribution. The magnetic field and the observation level of the WCD array remained unchanged in all simulations. The ground was placed at $5\,200\, {\rm m}$ above sea level, corresponding to the approximate altitude of some of the sites being considered for SWGO~\cite{SWGOFuture}.  
The Earth's magnetic field was fixed to the value at the ALMA site, in Chile.

\begin{figure}[ht]
	\centering
	\includegraphics[width = 0.9\linewidth]{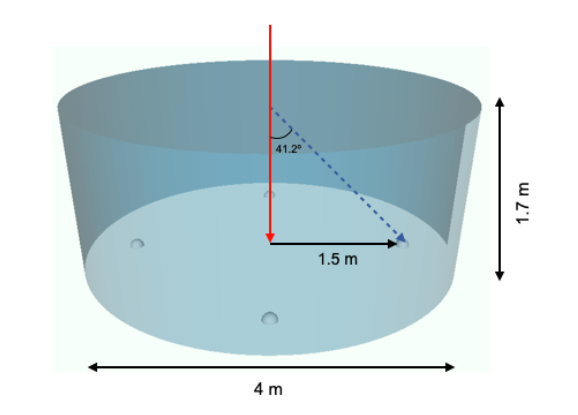}
	\caption{Sketch of the WCD unit employed in this study. The cylindrical tank is filled with water, and 4 PMTs (Photo Multiplier Tubes) are placed at the bottom of the structure. Taken from~\cite{wcd4pmt}.}
	\label{fig:WCDStation}
\end{figure}

The response of the WCD stations was emulated with a parameterisation of the signal as a function of particle energy obtained with the Geant4 toolkit~\cite{Geant4}. The signals induced by shower particles were obtained by injecting them at the centre of the detector in the vertical direction. 
A sketch of a WCD unit is shown in Fig.\,\ref{fig:WCDStation}. The single-layered WCD unit with multiple photo-sensors at the bottom is one of the candidate designs for the stations being considered for SWGO~\cite{wcd4pmt}. The parameterization of the average response of the WCD is obtained for electrons, muons and protons, representative of the electromagnetic, muonic and hadronic components of the shower, respectively.
It is important to note the discrimination shall be done through two shower quantities: \Smu and \Sem. As such, the lack of fluctuations on the parameterization, due to light collection and particle trajectories, would have an impact on the resolution of the reconstructed \Smu and \Sem. The impact of the experimental resolution on the reconstruction of these shower parameter will be discussed in~\ref{sec:reconstruction}.

With these simulations, we have computed \Sem and \Smu for each simulated neutrino and background proton shower at the ground array. The simulated values of \Sem and \Smu for signal and background events are fed into ROOT's Toolkit for Multivariate Data Analysis (TMVA)~\cite{TMVA} to separate the two classes of events as described in the next section.

%%%%%%%%%%%%%%%%%%%%%%%%%%%%%%%%%%%%%%%%%%%%%%%%%%%%%%%%%%%%%%%%%%%%%%%%%%
\section{Discriminating signal and background}
\label{sec:discrimination}

\begin{figure}[ht]
	\centering
	\includegraphics[width = 0.9\linewidth]{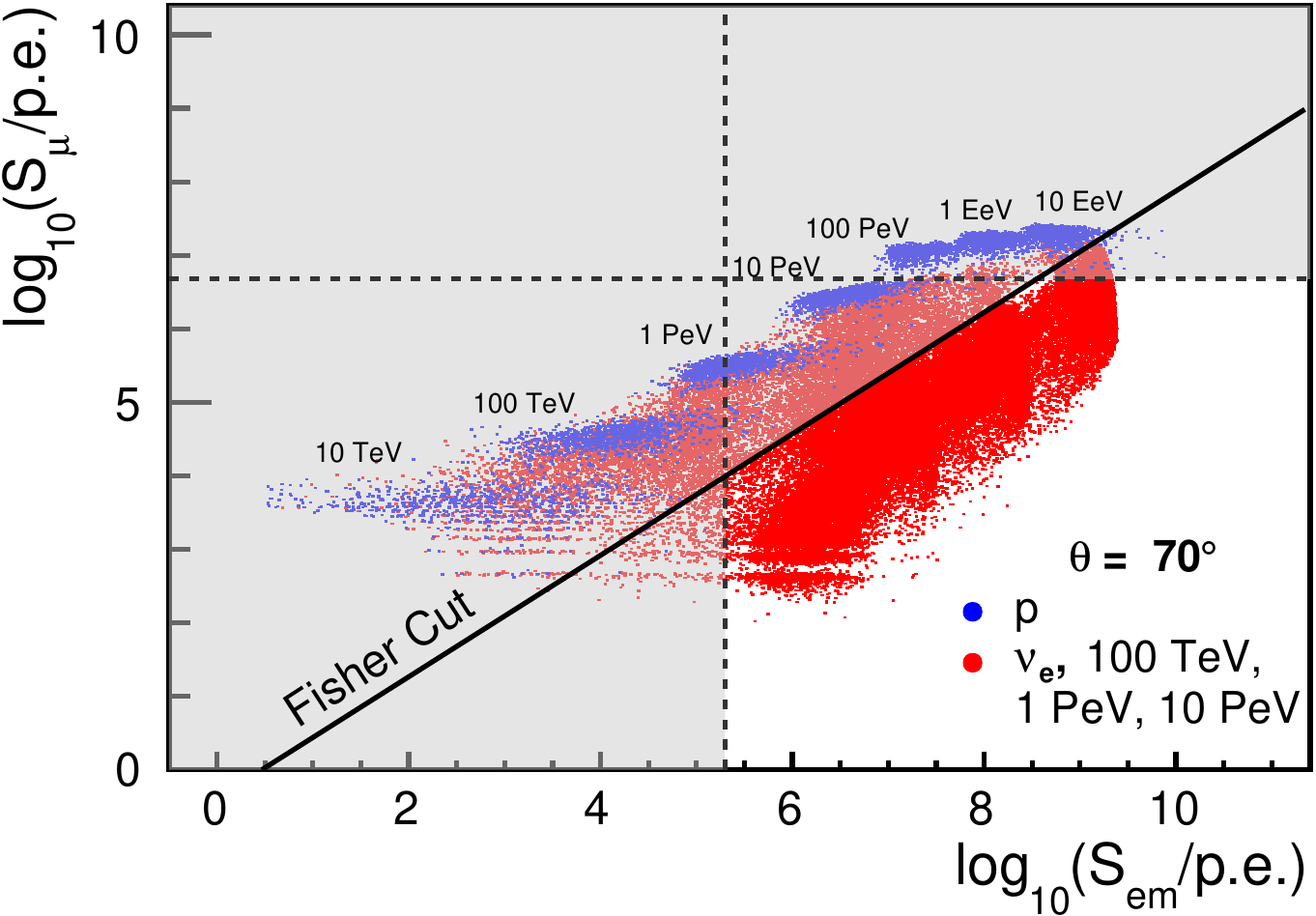}
	\caption{Fisher cut (solid line) applied in the discrimination between neutrino and proton-induced showers for $\theta=70^{\circ}$. Red dots represent neutrino events while blue dots represent proton-induced showers. The dotted vertical (horizontal) line corresponds to the cut in $\log_{10}S_\mathrm{em}$ ($\log_{10}S_\mu$) to reject all background proton events (see text for details). Only events that do not fall in the shaded gray region are considered as neutrino candidate events.}
	\label{fig:FishCut}
\end{figure}

The aim of this work was to minimise the background so that any neutrino candidate would be significant, at the expense of a smaller neutrino identification efficiency. This was achieved with a Fisher linear discriminant analysis performed in the parameter space of $\log_{10} (S_{\rm \mu})$ vs $\log_{10} (S_{\rm em})$. The cut in the Fisher discriminant is derived independently for each simulated zenith angle considering all the simulated proton energies ($10\,$TeV-$10\,$EeV) and neutrinos with fixed energy from $100\,$TeV to $10\,$PeV. An example is shown in Fig.~\ref{fig:FishCut} for the case of $\theta=70^\circ$. It was found that the optimal Fisher cut varies with the zenith angle, but not with the primary energy.

Two additional cuts were introduced to achieve a background-free discrimination. 
Neutrino events have Fisher values predominantly above $\sim 0.5$. However, also a small fraction of low-energy proton events typically characterised by small values of \Sem can fulfil the Fisher cut. For all values of zenith angle, a cut on $\log_{10}(S_{\rm em}/{\rm p.e.}) > 5.3$ removes the majority of these background events, while minimising the loss of neutrino events. An example is shown in Fig.\,\ref{fig:FishCut}.

A second, zenith-dependent cut on \Smu was introduced to remove the contamination due to the highest-energy proton background showers. Cascades induced by protons with energies above $1\,{\rm PeV}$ produce larger muonic signals than those induced by neutrinos with energies in the $100\,{\rm TeV}\,-\,10\,{\rm PeV}$ range. By limiting the maximum value of \Smu, these background events are eliminated with minimal loss of neutrino events as can be seen in the example in Fig.\,\ref{fig:FishCut}.

Within the squared region defined by the \Sem and \Smu cuts, the value of the Fisher cut can be further adjusted to remove all background events.

%%%%%%%%%%%%%%%%%%%%%%%%%%%%%%%%%%%%%%%%%%%%%%%%%%%%%%%%%%%%%%%%%%%%%%%%%%
\section{Sensitivity of a ground array to neutrinos}
\label{sec:method}

% Event Rate Formula 
To estimate the sensitivity of a gamma-ray ground-based observatory to neutrinos we have calculated the expected neutrino event rate $\d N_\nu/\d t$ given by the following equation,
\begin{eqnarray}
    \frac{\d N_\nu}{\d t} &=& \int^{E_{\nu,{\rm max}}}_{E_{\nu,{\rm min}}} \frac{\d \Phi_\nu}{\d E_\nu}(E_\nu) \, \frac{1}{m} \, \sigma(E_\nu) \, M_{\rm eff}(E_\nu) \, \d E_{\nu} \,,
    \label{eq:Sensitivity}
\end{eqnarray}
where $\d \Phi_\nu/\d E_\nu$ denotes the differential flux of incoming neutrinos, $m$ is the mass of an air nucleon and $\sigma(E_\nu)$ is the neutrino-nucleon cross section. $M_{\rm eff}(E_\nu)$ is the effective mass of the detector (see below), while $E_{\nu,{\rm min}}$ and $E_{\nu,{\rm max}}$ denote the integration limits used for the sensitivity calculation. 

In this Section we study the sensitivity to electron neutrinos only. The sensitivity to all neutrino flavors will be addressed in Section~\ref{sec:resultsall}.

%-------------------------------------------------------------------
\subsection{Electron Neutrino Flux}

An astrophysical flux of VHE electron neutrinos and anti-neutrinos was measured at the IceCube neutrino observatory up to a few PeV~\cite{IceCubeFlux}. The flux 
of $\nu_{\rm e}$ and $\bar\nu_{\rm e}$ can be approximated by: 
\begin{equation}
    \frac{\d \Phi_\nu}{\d E_\nu}(E_\nu) = k^\prime \left( \frac{E_\nu}{E_0} \right)^{-2.53},
    \label{eq:NuFlux}
\end{equation}
where $E_0=10^5\,{\rm GeV}$, and $k^\prime= k E_0^{-2.53} \equiv 4.98 \times 10^{-18}\,{\rm GeV^{-1}\,cm^{-2}\,s^{-1}\,sr^{-1}}$. In this work, we discuss the detection of neutrinos with energy above $100\,$TeV, where the flux of astrophysical neutrinos dominates over the one by atmospheric neutrinos. As such, we will use for electron-neutrinos the flux given in Eq.\,(\ref{eq:NuFlux}) reduced by a factor of two, assuming an equal content of $\nu_e$ and $\bar\nu_e$ at Earth. Moreover, as in this work we intend only to have an estimate of the number of neutrinos that could be detected by a generic gamma-ray observatory through the use of inclined showers, we consider only the mean values reported by IceCube, i.e., we neglect for the up-coming calculations the experimental errors claimed by the experimented.

%-------------------------------------------------------------------
\subsection{Neutrino-nucleon Cross-section}

\begin{figure}[ht]
	\centering
	\includegraphics[width = 0.9\linewidth]{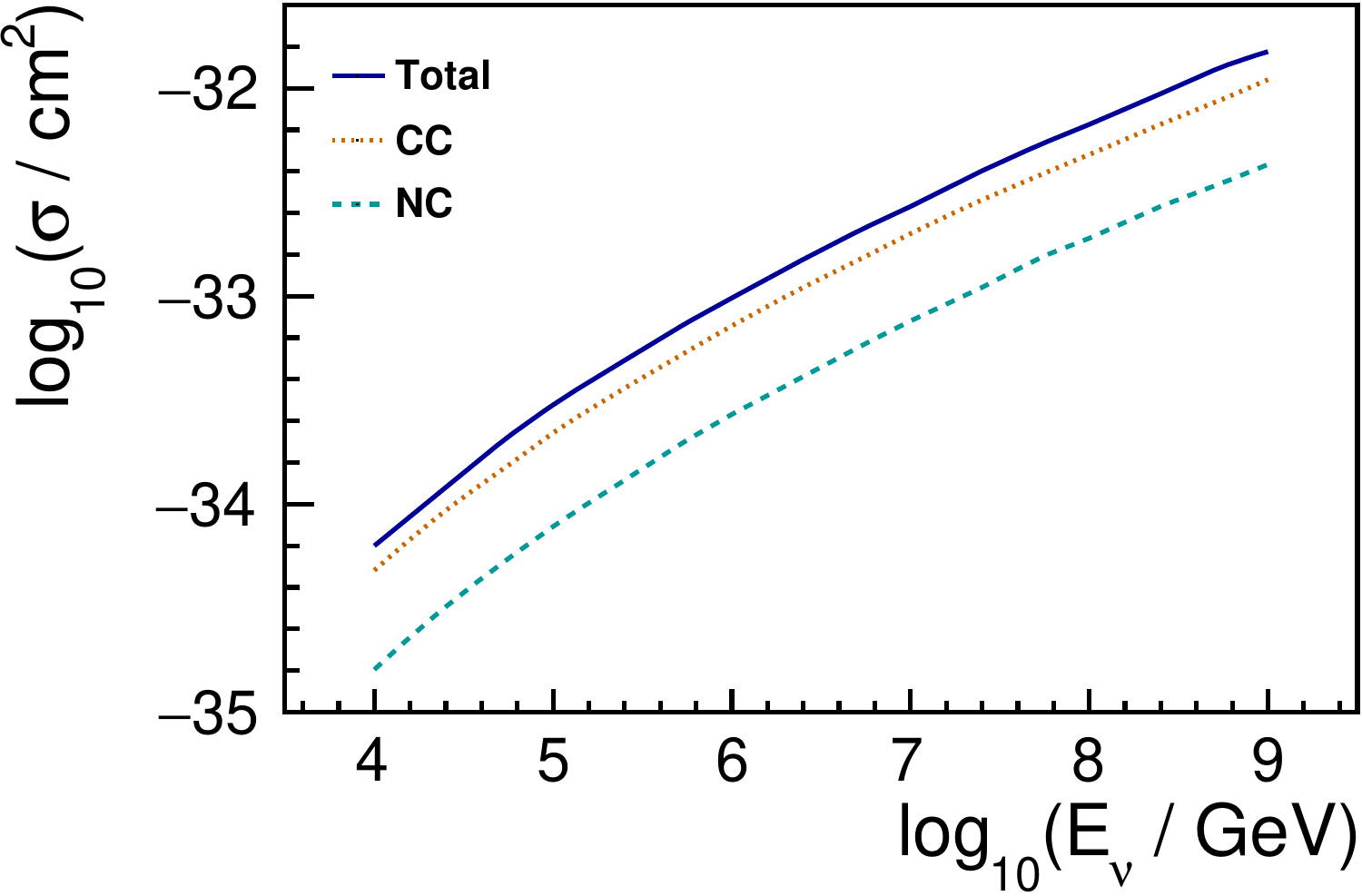}
	\caption{Neutrino-nucleon charged current (CC), neutral current (NC), and total (CC+NC) cross sections as a function of neutrino energy $E_\nu$. Values taken from~\cite{NuXSecs}.}
	\label{fig:NuNucXSection}
\end{figure}

In Eq.\,(\ref{eq:NuFlux}) we use the values of the neutrino-nucleon cross-section as a function of energy from~\cite{NuXSecs}, distinguishing  between charged current (CC) and neutral current (NC) neutrino interactions, as shown in Fig.\,\ref{fig:NuNucXSection}. 

%-------------------------------------------------------------------
\subsection{Neutrino efficiency and effective mass}

The effective mass represents the amount of matter within which an interacting neutrino can be identified. Eq.\,(\ref{eq:effMass}) gives the effective mass as a function of the zenith angle $\theta$, and the energy of the incoming neutrino $E_\nu$:
\begin{equation}
M_{\rm eff}^\theta(E_\nu,\theta)= 2\pi A  \sin \theta \cos\theta \hspace{1mm}\int_D\varepsilon_\nu(E_\nu, \theta, D) \hspace{1mm} dD\ . % \qquad [{\rm g} ].
\label{eq:effMass}
\end{equation}
The function $\varepsilon_\nu(\theta, D, E_{\nu})$ denotes the probability of identifying a neutrino considering the cuts introduced in Section~\ref{sec:discrimination}. It is a function of the slant depth of the neutrino point of first interaction of the neutrino, $D$, (expressed in ${\rm g\,cm^{-2}}$ and measured from ground), the energy of the neutrino $E_\nu$ (given in GeV), and the angle of incidence $\theta$ (in ${\rm radians}$). The surface area of the array is denoted as $A$, and was fixed at a value $A=1\, {\rm km}^2$.

\begin{figure}[ht]
	\centering
	\includegraphics[width = 0.9\linewidth]{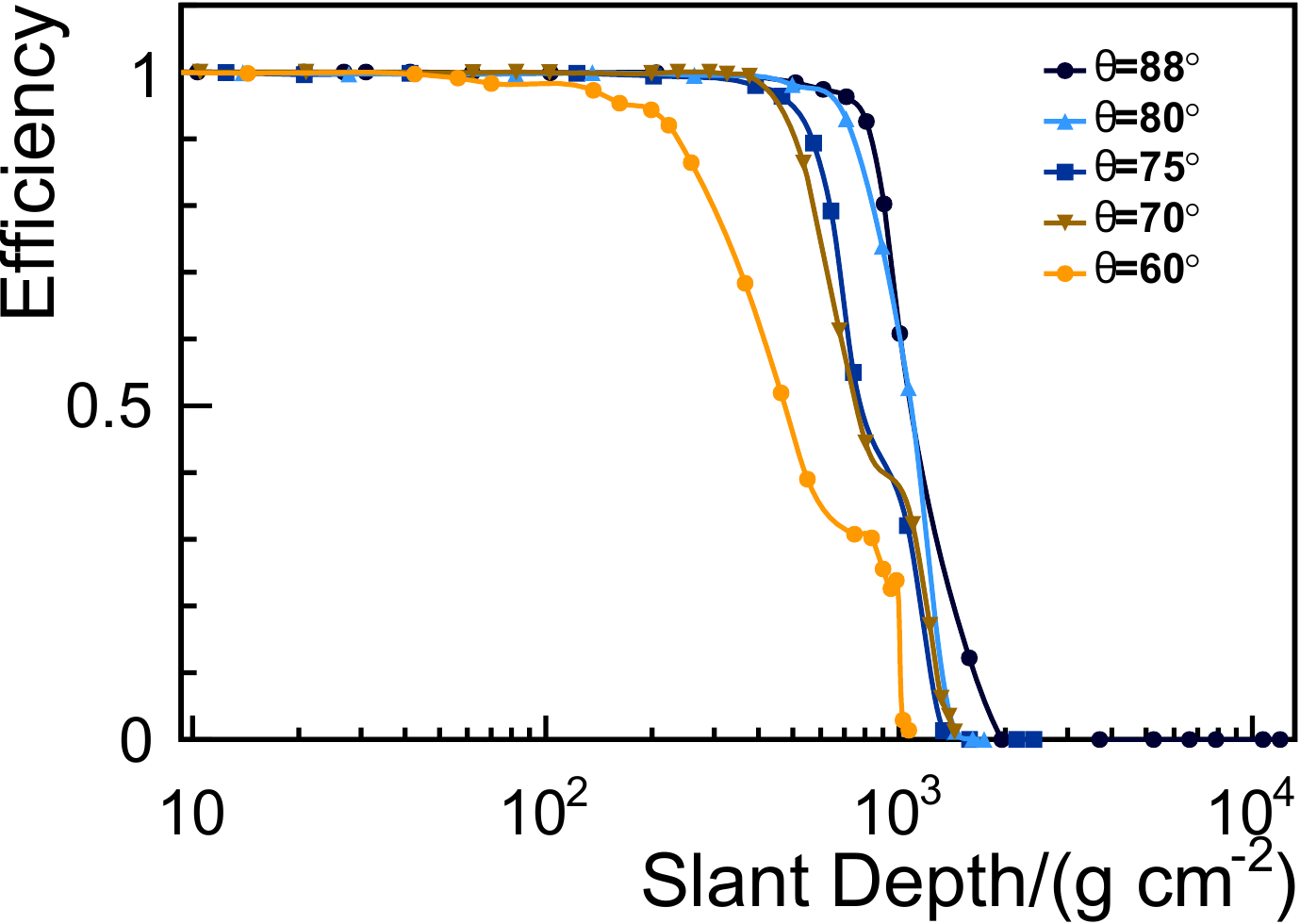}
	\caption{Neutrino identification efficiency as a function of neutrino interaction slant depth (measured from ground), for simulated neutrino-induced of $E_\nu=1\,{\rm PeV}$, and $\theta= 60^{\circ}$, $70^{\circ}$, $75^{\circ}$, $80^{\circ}$ and $88^{\circ}$.}
	\label{fig:effCurves}
\end{figure}

The neutrino identification efficiency $\varepsilon_\nu(E_\nu,\theta,D)$ is obtained as the ratio of the number of neutrino points within the area delimited by the cuts (white region in Fig.\,\ref{fig:FishCut}) and the total number of simulated neutrino points for a given zenith angle, energy and interaction depth. An example is depicted in Fig.\,\ref{fig:effCurves} for $E_\nu=1\,{\rm PeV}$ and several zenith angles as a function of $D$. As expected the neutrino identification efficiency decreases for showers initiated far from the ground since those are more similar to showers induced by protons that typically interact in the upper layers of the atmosphere. 

For each primary neutrino energy, five values of $\theta$ are considered: $60^\circ$, $70^\circ$, $75^\circ$, $80^\circ$ and $88^\circ$. The integration in $D$ of Eq.\,(\ref{eq:effMass}) is done using a cubic spline interpolation to the discrete values of $\varepsilon_\nu(E_\nu,\theta, D)$\footnote{$E_\nu$ and $\theta$ are fixed for each case.}. This results in the effective mass values for each value of $\theta$ reported in Table~\ref{tab:EffMass}.
   
\begin{table}[ht]
	\centering
	\renewcommand{\arraystretch}{1.3}
	\begin{tabular}{c | c}\hline
		\textbf{$\theta$}		& \textbf{$M_{\rm eff}^\theta(E_\nu=1{~\rm PeV},\theta)\,[{\rm g}]$} \\
		\hline
		$60^\circ$			& $9.73\times 10^{12}$							\\
		$70^\circ$			& $1.27\times 10^{13}$							\\
		$75^\circ$			& $1.65\times 10^{13}$						\\
		$80^\circ$			& $9.09\times 10^{12}$						\\
		$88^\circ$			& $2.21\times 10^{12}$						\\
		\hline
	\end{tabular}
	\caption{Effective mass as given in Eq.\,(\ref{eq:effMass}) for neutrino-induced showers with $E_\nu=1~{\rm PeV}$ and several values of $\theta$.}
	\label{tab:EffMass}
\end{table}

% Comment on the evolution of the effective mass values with theta?

The total effective mass for a given neutrino energy is obtained by integrating the effective mass in zenith angle, $\theta \in [60^\circ ; 89^\circ]$.
The integration in zenith angle is achieved by applying a cubic spline interpolation to the $M_{\rm eff}^\theta(\theta, E_{\nu})$ values listed in Table\,\ref{tab:EffMass} for the case of $E_\nu = 1\, {\rm PeV}$. This yields a total effective mass for the reference energy $E_\nu=1\,{\rm PeV}$ of $M_{\rm eff}\simeq 2.97\times 10^{14}\,{\rm g\, sr}$.

%-------------------------------------------------------------------
\subsection{Electron Neutrino Interactions}
%Discuss Neutral and Charged current interactions

\begin{figure}[ht]
	\centering
	\includegraphics[width = 0.9\linewidth]{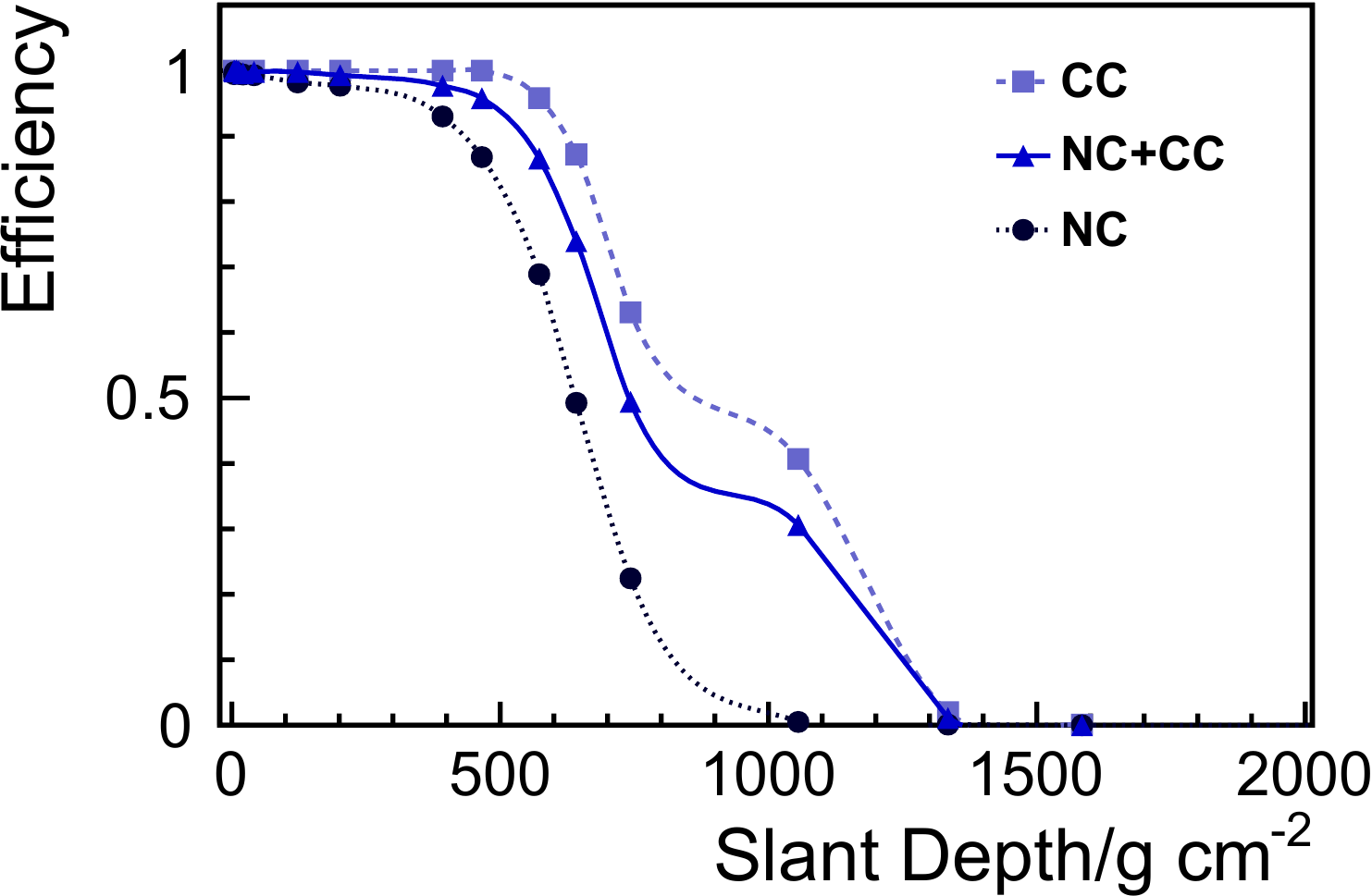}
	\caption{Neutrino identification efficiency obtained for showers induced by $1\,{\rm PeV}$ neutrinos with $\theta=80^\circ$. Interactions are either selected at random between CC and NC according to their relative weight in the total cross section (curve labelled as NC+CC), or set to only CC or only NC interactions.}
	\label{fig:NCCCEff80}
\end{figure}

The neutrino detection efficiency and the effective mass depend on the neutrino interaction channel. In Fig.\,\ref{fig:effCurves} and Table\,\ref{tab:EffMass}, the interaction channel, either CC or NC, was randomly chosen according to their relative weights in the total cross-section. However, in CORSIKA simulations, the interaction can be chosen so that neutrinos only interact via CC or NC, allowing the estimation of the sensitivity for each interaction channel. An example of the resulting neutrino identification efficiency is presented in Fig.\,\ref{fig:NCCCEff80}, for $E_\nu=1\,{\rm PeV}$ and $\theta=80^\circ$.

As seen in Fig.\,\ref{fig:NCCCEff80}, the electron neutrino identification efficiency considering only CC interactions has non-zero values at a larger distance from ground than the one obtained using only NC interactions. This happens because, in CC interactions, the total energy of the $\nu_e$ is transferred to an electromagnetic shower, from the energetic electron produced in the interaction, and a hadronic shower from the collision with the nucleon of the atmosphere.

In NC interactions, instead of an electron, an electron-neutrino will be produced. Hence, only the typically less energetic hadronic shower can be detected reducing the efficiency. In Fig.\,\ref{fig:NCCCEff80}, it is also shown the more realistic case of the efficiency when CC and NC interactions are chosen at random depending on their relative weight in the total neutrino-nucleon cross section.
As expected, the curve NC+CC is in between the CC and NC curves.

Integrating Eq.\,(\ref{eq:effMass}) in zenith angle for a fixed energy, yields the 
effective masses reported in Table\,\ref{tab:EffMasses} for \(\x{E_\nu=1\,{\rm PeV}}\). 
\begin{table}[ht]
	\centering
	\renewcommand{\arraystretch}{1.5}
	\begin{tabular}{c | c}\hline
		Interaction 		& $M_{\rm eff}(E_\nu=1\,{\rm PeV})\,[{\rm g\,sr}]$ \\
		\hline
		CC			& $3.60\times 10^{14}$							\\
		NC		    & $2.27\times 10^{14}$							\\
		Total		& $2.97\times 10^{14}$						\\
		\hline
	\end{tabular}
	\caption{Effective mass for the different neutrino interaction channels CC and NC, with $E_\nu=1~{\rm PeV}$. Total corresponds to the case where CC or NC are chosen randomly.}
	\label{tab:EffMasses}
\end{table}

%Eq.\,(\ref{eq:Sensitivity}) can be integrated in energy ($1-2\,$PeV) 
%for every interaction channel configuration, yielding the rates listed in Table \ref{tab:NC_CC_Sensitivities}. 
%The event rate obtained when the interaction type is not fixed, $3.12 \times 10^{-2}\, {\rm yr^{-1}}$, exceeds the sum of the individual cases where the first interaction occurs either exclusively via CC or NC, $2.54\times 10^{-2} \, {\rm yr^{-1}}$. Thus, the approach in which CC or NC are chosen according to their relative cross-sections maximises the number of expected events per year. 

%\begin{table}[ht]
%	\centering
%    \renewcommand{\arraystretch}{1.5}
%	
%	\begin{tabular}{c | c}\hline
%		Interaction		& $\frac{dN}{dt}[{\rm yr^{-1}}]$ \\
%		\hline
%		CC		 & $2.29\times 10^{-2}$							\\
%		NC		 & $2.50 \times 10^{-3}$							\\
%		Total    &  $3.12 \times 10^{-2}$ \\ 
%		%Total ($M_{\rm eff}$ given by Eq.\ref{eq:effMassVsE})	 & $3.695 \times %10^{-2}$						\\
%		\hline
%	\end{tabular}
%	\caption{Sensitivity of a wide-field ground-based gamma-ray observatory to astrophysical neutrinos, according to the type of the first interaction. $E_\nu$ spanning from $1{\rm PeV}$ to $2\,{\rm PeV}$}
%	\label{tab:NC_CC_Sensitivities}
%\end{table}

%%%%%%%%%%%%%%%%%%%%%%%%%%%%%%%%%%%%%%%%%%%%%%%%%%%%%%%%%%%%%%%%%%%%%%
\section{Sensitivity to downward-going $\nu_e$}
\label{sec:resultsdown}

Eq.\,(\ref{eq:Sensitivity}) can be integrated over energy to obtain the electron neutrino event rate. 

This is achieved by applying a cubic spline interpolation to estimate the effective mass values for neutrino energies between $100\,{\rm TeV}$ and $10\,{\rm PeV}$. The effective mass for energies outside this range is approximated via extrapolation.
The estimated electron neutrino event rates are given in Table\,\ref{tab:Sensitivities}. Different values of $E_{\nu,{\rm min}}$ and $E_{\nu,{\rm max}}$ were used in Eq.\,(\ref{eq:Sensitivity}) to study the dependence of the event rate on both the minimum energy above which the flux can be considered to be purely astrophysical with a negligible contamination from atmospheric neutrinos, and on the maximum energy to which the astrophysical flux could extend without a cutoff. As can be seen in Table\,\ref{tab:Sensitivities}, a rate of 0.3 electron neutrinos per year can be detected.

\begin{table}[ht]
	\centering
	\renewcommand{\arraystretch}{1.5}
	\begin{tabular}{c | c}\hline
		$E_{\nu,{\rm min}} - E_{\nu,{\rm max}}$		& $\frac{dN}{dt}(E_{\nu})[{\rm yr^{-1}}]$ \\
		\hline
%		$100\,{\rm TeV}-200\,{\rm TeV}$	& $3.82\times 10^{-2}$		\\
		$100\,{\rm TeV}-1\,{\rm PeV}$   & $1.30 \times 10^{-1}$ \\ 
		$100\,{\rm TeV}-10\,{\rm PeV}$  & $2.06\times 10^{-1}$	 \\
		$100\,{\rm TeV}-100\,{\rm PeV}$ & $3.01 \times 10^{-1}$	 \\ \hline
%		$1\,{\rm PeV}-2\,{\rm PeV}$	    & $3.12\times 10^{-2}$		\\
		$1\,{\rm PeV}-10\,{\rm PeV}$	& $1.06 \times 10^{-1}$		\\
		$1\,{\rm PeV}-100\,{\rm PeV}$	& $1.72 \times 10^{-1}$		\\ 

		\hline
	\end{tabular}
    \caption{Even rate, given by Eq.\,(\ref{eq:Sensitivity}), for electron neutrinos only in a wide-field ground-based gamma-ray observatory ($A=1\,{\rm km^2}$), for different ranges of $E_\nu$. The rates are obtained in different ranges of $E_{\nu,{\mathrm min}}$ and $E_{\nu,{\mathrm max}}$ in Eq.\,(\ref{eq:Sensitivity}).
    }
	\label{tab:Sensitivities}
\end{table}

\begin{figure}[ht]
	\centering
	\includegraphics[width = 0.9\linewidth]{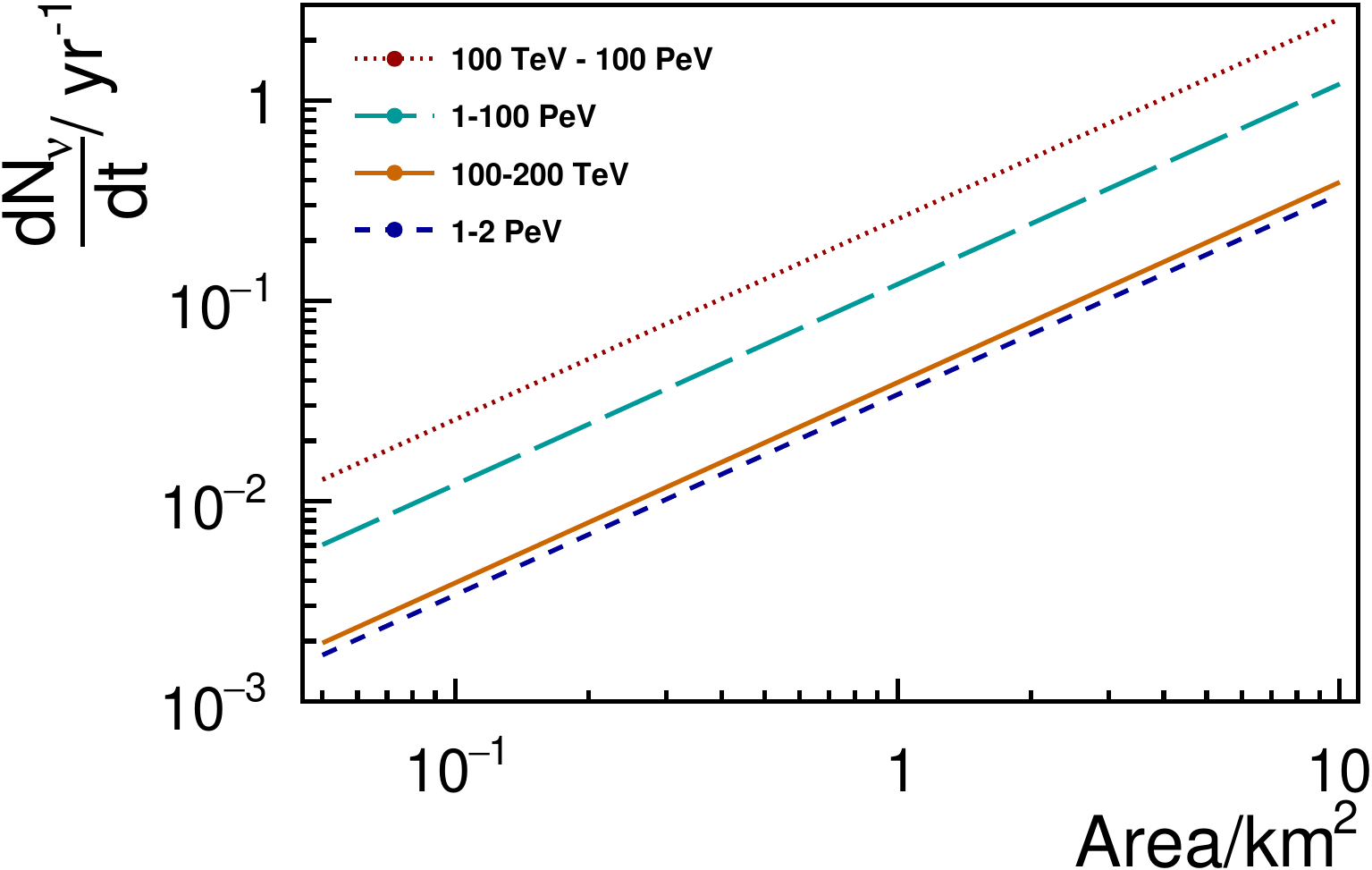}
	\caption{Number of electron neutrinos expected to be detected and identified per year as a function of the area of the detector. Three curves are presented corresponding to different values of $E_{\nu,{\mathrm min}}$ and $E_{\nu,{\mathrm max}}$, ranging from $1\,{\rm PeV}$ to $2\,{\rm PeV}$ and $100\,{\rm PeV}$, as well as from $100\,{\rm TeV}$ to $200\,{\rm TeV}$ and $100\,{\rm PeV}$. For reference the LHAASO ground array has an area of $A=1\,{\rm km^2}$.}
	\label{fig:NusPerYear}
\end{figure}

The estimates of sensitivity given in Table\,\ref{tab:Sensitivities} can be extrapolated linearly to other values of detector surface area $A$. In Fig.\,\ref{fig:NusPerYear} we depict the electron neutrino event rates for as a function of $A$ for different values of $E_{\nu,{\mathrm min}}$ and $E_{\nu,{\mathrm max}}$.

% Describe procedure followed in the 1 PeV case -- Up to Total Effective Mass Calculation
% Mention repetition of procedure for  100 TeV and 10 PeV
% Effective Mass of Energies between 100 TeV and 10 PeV can be obtained via interpolation (cubic spline interpolant). Energies above 10 PeV require extrapolation
% Sensitivity calculation repeated for several energy ranges

%-------------------------------------------------------------------
\subsection{Impact of the array fill factor}

\begin{figure}[ht]
	\centering
	\includegraphics[width = 0.9\linewidth]{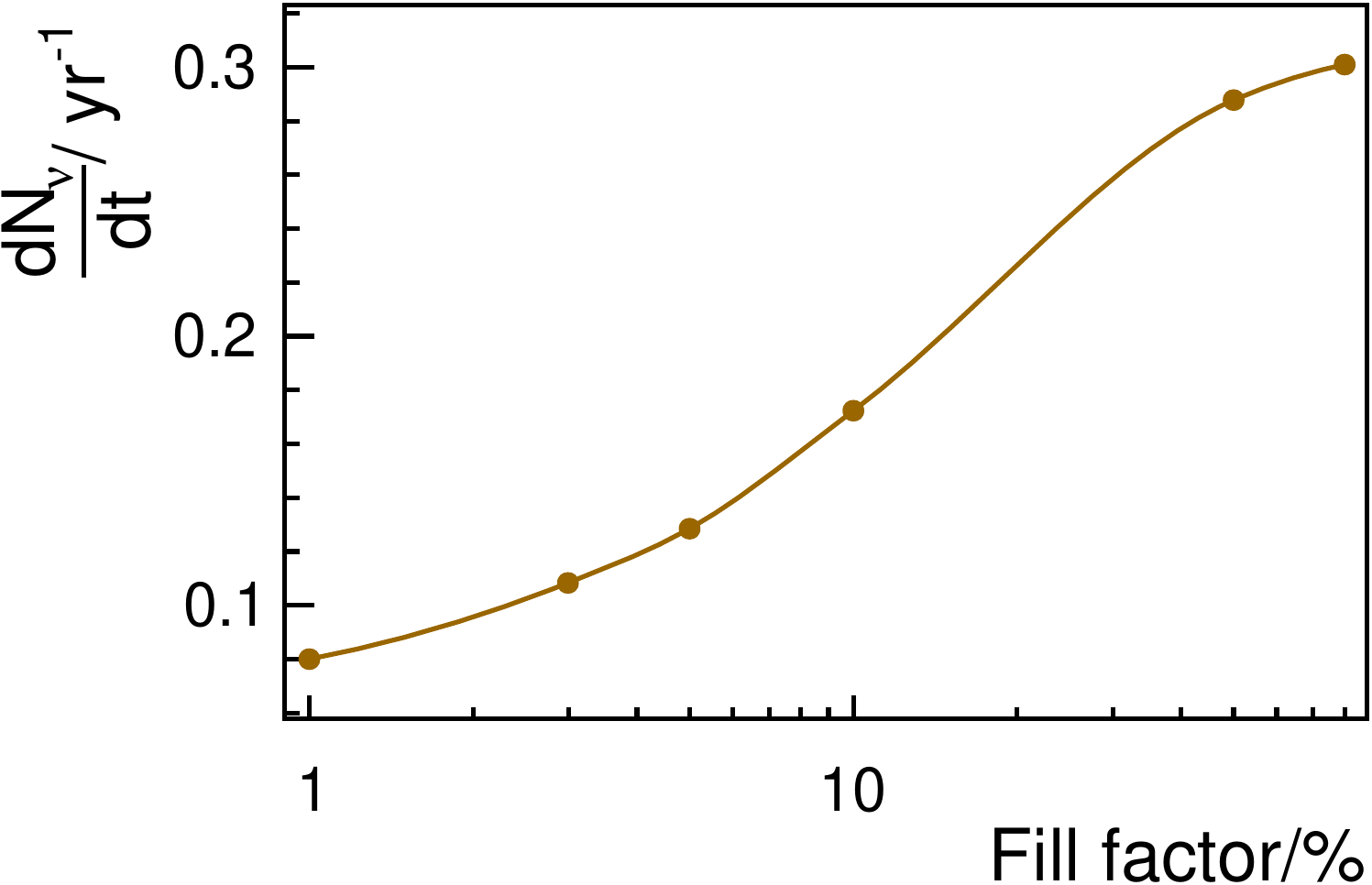}
	\caption{Estimated electron neutrino event rate as a function of the fill factor of the WCD array, see text for details. The event rate is obtained with Eq.\,(\ref{eq:Sensitivity}) for $E_{\nu,{\mathrm min}}=100$ TeV and $E_{\nu,{\mathrm max}}=100$ PeV, for an array with $A = 1\,{\rm km^2}$.}
	\label{fig:NusVsFF}
\end{figure}

The fill factor is defined as the ratio between the sum of the area of individual detectors and the total area of the array $A$.  
To infer the impact of the fill factor on the event rate, the procedure described previously is applied to a detector array of equal surface area ($1\,{\rm km^2}$) and variable fill factor. In this work we have  studied the sensitivity for fill factors of $1, 3, 5, 50$ and $80\%$, yielding the results in Fig.\,\ref{fig:NusVsFF}. All the cuts described in Section~\ref{sec:discrimination} were recomputed to ensure that all the simulated proton background events are rejected.

Taking as a reference LHAASO's fill factor of $4\%$~\cite{LHAASOLayout}, the estimated neutrino event rate decreases by a factor of $\approx 3$ when compared to the initially assumed $80\%$ fill factor. It is interesting to see that the event rate increases rather slowly for fill factors between 1\% and $\sim 5\%$ and more rapidly between $\sim 10\%$ and $\sim 50\%$.

%-------------------------------------------------------------------
\subsection{Impact of experimental resolution}
\label{sec:reconstruction}

We have also studied the impact of experimental resolution on the expected event rate. Gaussian smearings, denoted as $\sigma_{S_{\rm em}}$ and $\sigma_{S_{\rm \mu}}$, were applied to both electromagnetic (\Sem) and muonic (\Smu) signals of the neutrino and background events, respectively. 

\begin{figure}[ht]
	\centering
	\includegraphics[width=0.9\linewidth]{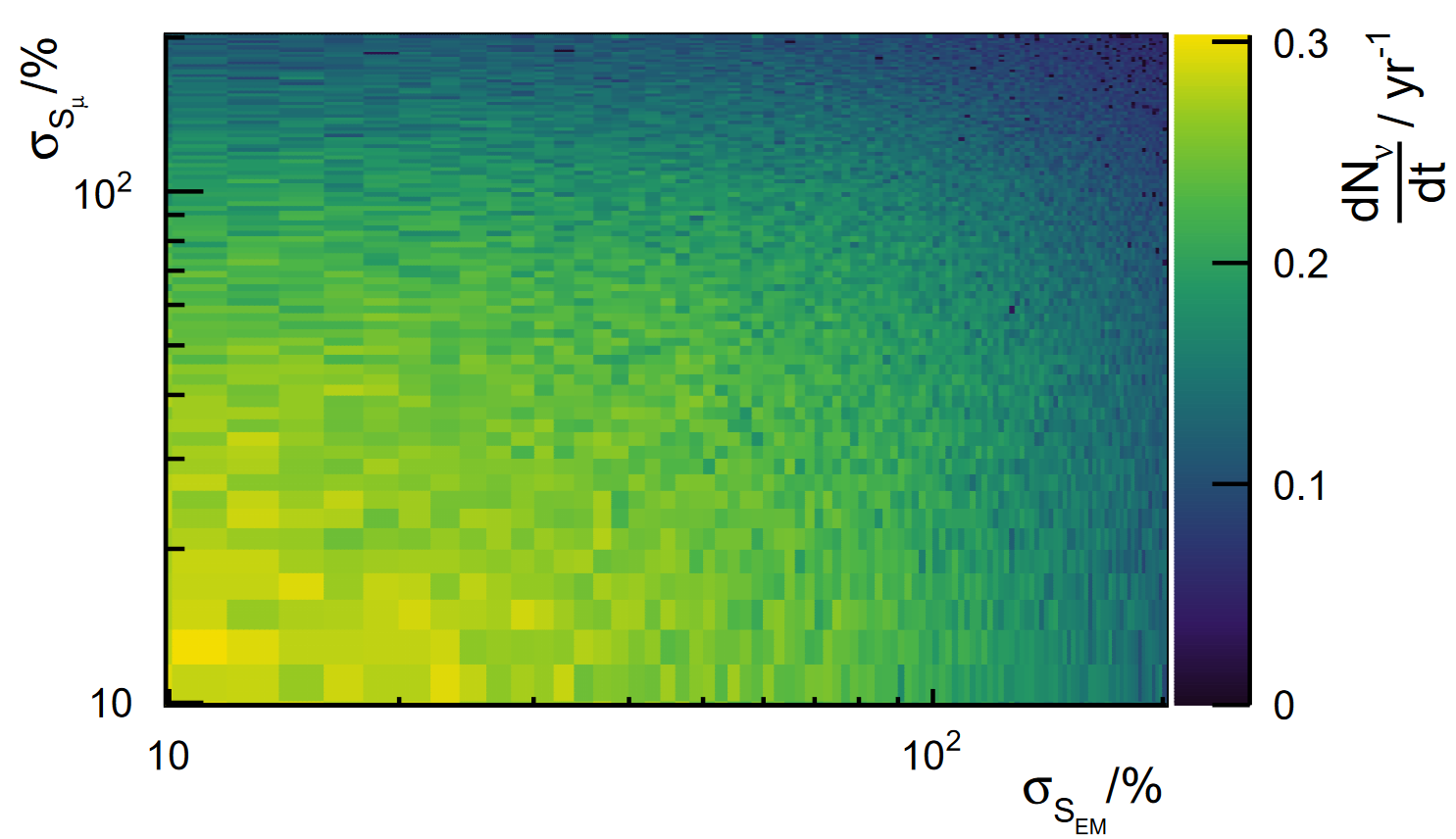}
	\caption{Electron neutrino event rate as a function of the experimental resolution on the discriminating variables \Sem and \Smu assumed to follow a Gaussian distribution of width $\sigma_{S_\mu}$ and $\sigma_{S_{\rm em}}$. The rate was obtained with Eq.\,(\ref{eq:Sensitivity}) for the range of energies $E_{\nu,{\mathrm min}}=100\,{\rm TeV}$ and $E_{\nu,{\mathrm max}}=100\,{\rm PeV}$, assuming a detector surface area $A=1\,{\rm km^2}$.}
	\label{fig:SmearVsNusLargeScale}
\end{figure}

After applying the smearing, the previously derived cuts on the Fisher discrimination described in Section~\ref{sec:discrimination} were recomputed to ensure that all simulated background events are rejected. Assuming again an array area of $A=1\,{\rm km}^2$ with an 80\% fill factor, the resulting neutrino event rates are presented in Fig.\,\ref{fig:SmearVsNusLargeScale}. Larger values of $\sigma_{S_\mu}$ and/or $\sigma_{S_{\rm em}}$ result in progressively lower event rates and hence lower sensitivity, as would be expected. Degradation of the expected number of neutrinos by a factor of $2$ is only achieved when the smear applied to the electromagnetic or muonic signal reaches an extreme value of about $200\%$. However, at PeV energy, the reconstruction resolutions of \Sem and \Smu are expected to be a few tens of percent. The reduced impact on the event rate reflects the robustness of this methodology to a possible degradation of the signal due to reconstruction.

The ability to reconstruct the geometry (arrival direction and core position) of the neutrino-induced shower events, was also investigated using a simple reconstruction algorithm. The reconstruction is performed by fitting the arrival times of the first particles reaching each WCD station to a conic shower front. The curvature of the front was taken from\cite{conic_geom}, without any further optimisation. This test was done considering an array of $A=1\,{\rm km^2}$ and a fill factor of $5\%$.

\begin{figure}[ht]
	\centering
	\includegraphics[width=0.9\linewidth]{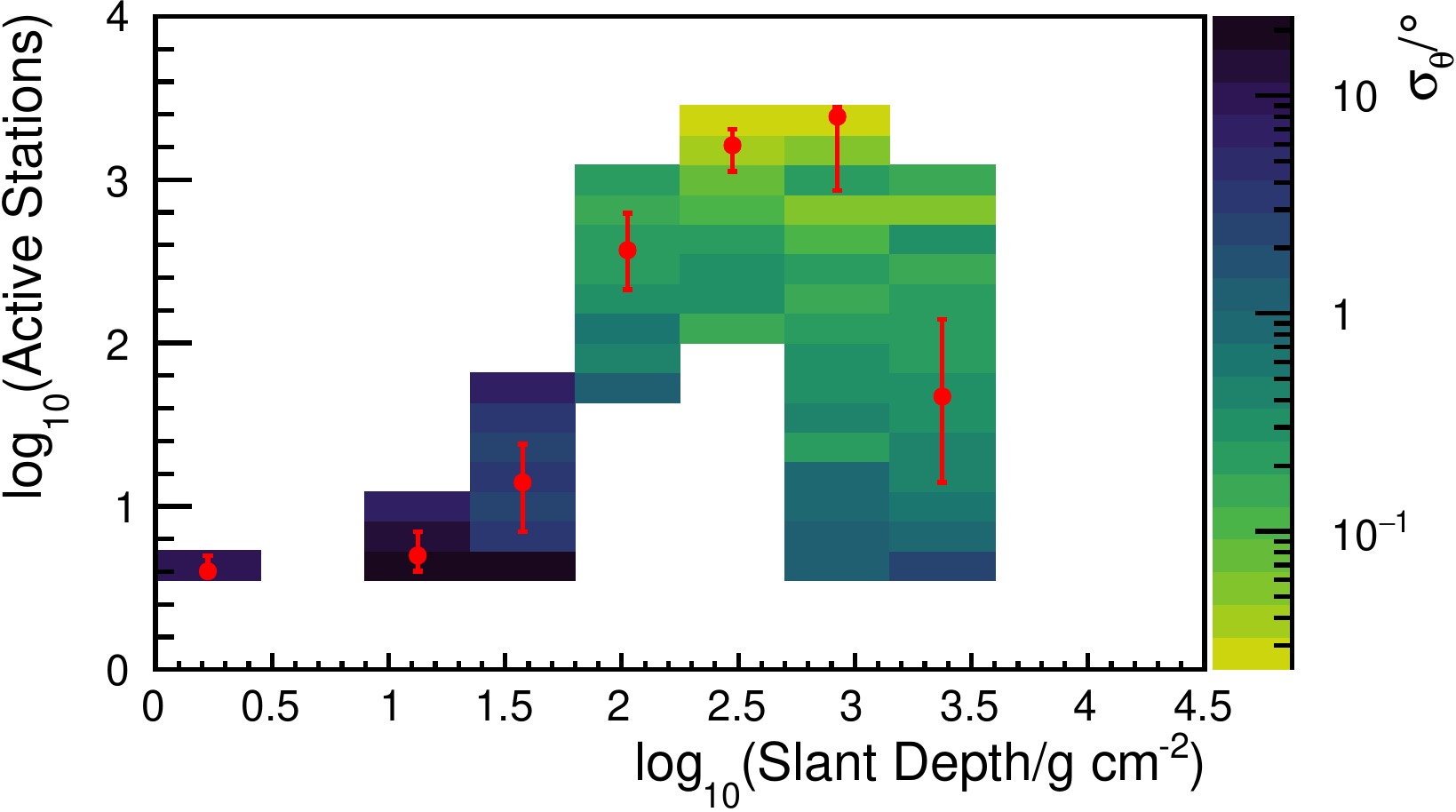}
	\caption{Angular reconstruction resolution as a function of the neutrino interaction slant depth (measured from ground), and of the number of active stations, for neutrinos with $E_\nu = 1\,{\rm PeV}$ and $\theta=75^\circ$. Red points denote the median, and the error bars the standard deviation of the event distribution within each slant depth bin.}
	\label{fig:XNstationSigmaTheta}
\end{figure}

In figure~\ref{fig:XNstationSigmaTheta}, we show a density plot for the angular reconstruction resolution, $\sigma_\theta$, as a function of the neutrino interaction slant depth and the number of active stations. The resolution $\sigma_\theta$ is defined as the $68\%$ containment of the difference between the simulated and reconstructed angle. From this figure, it can be seen that the precision of the shower axis reconstruction depends both on the distance of the neutrino interaction point to the ground, and on the number of triggered stations. If the interaction happens close to the ground, the shower footprint is small, leading to a poor reconstruction. However, if the interaction happens at $\gtrsim 100\,{\rm g\,cm^{-2}}$ it is possible to achieve angular resolutions better than $\sim 1^\circ$. 

Experimentally, one could apply a cut on the number of active stations. For instance, it was seen that requiring at least $\sim 30$ active stations would allow having a reconstruction resolution better than $\sim 5^\circ$. The introduction of such a condition would lead to a small $\sim 10\%$ decrease in the neutrino identification efficiency and effective mass, resulting in a proportionately lower neutrino event rate.

In figure~\ref{fig:XNstationSigmaTheta}, it can also be seen that for showers with large slant depths ($D\gtrsim 1000\,{\rm g\,cm^{-2}}$), the number of active stations can have significant variations being intrinsically connected to the shower development. However, the plot also displays the median and the standard deviation of the number of events, evidencing that most of the showers will lead to a large number of active stations. It should also be pointed out that while the number of active stations affects the quality of the reconstruction, better resolutions can be attained for neutrino-induced showers that interact higher in the atmosphere. This happens because even though fewer particles are reaching the ground, the shower footprint is more extended due to the longer shower development through the atmosphere, easing the reconstruction of the geometry.

It was verified that the order of magnitude of the claimed geometric reconstruction resolution is the same for all the energies and angles considered in this work.

Finally, it is important to note that the provided values on the reconstruction resolutions should be taken as upper limits. Dedicated reconstructions of inclined showers are expected to improve the angular resolution~\cite{AugerInclined}.

%----------------------------------------------------------------
\subsection{Impact of the limited simulation statistics}

The flux of background proton-induced showers greatly exceeds the expected flux of neutrinos, implying that a reliable observation of neutrino events requires a large background rejection factor. 
Simulations are needed to establish the cuts and assess a possible contamination by proton showers in order to get a significant detection in case a neutrino candidate is observed. However, the available simulations are limited in statistics due to limited computational resources and computing time.

\begin{figure}[ht]
	\centering
	\includegraphics[width=0.9\linewidth]{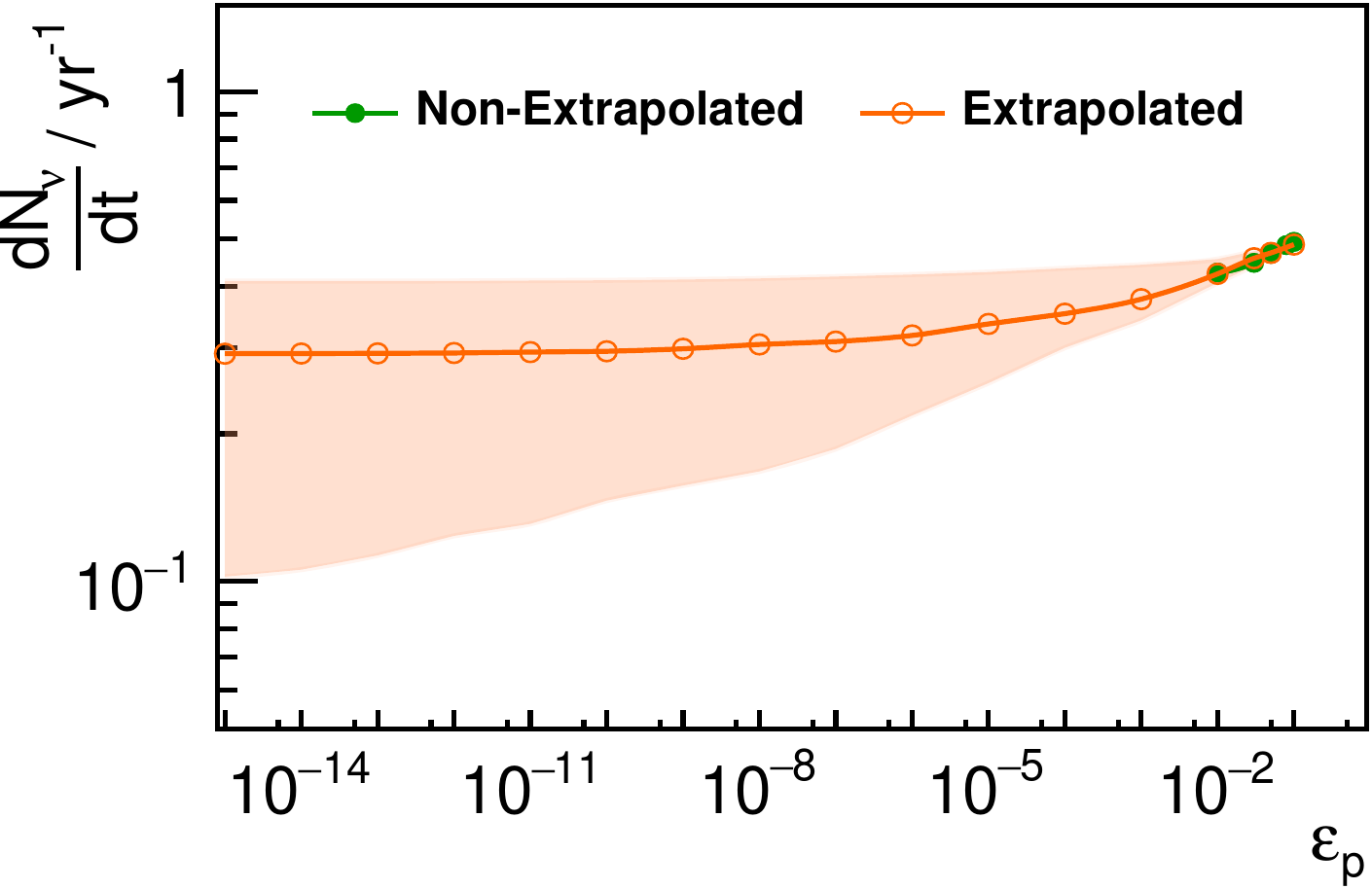}
	\caption{Neutrino event rate as a function of proton contamination fraction $\varepsilon_{p}$, using only simulated data (green dots with line), and extrapolating from available data points (orange dots with line and band).}
	\label{fig:EventRatesVsEpsilonP}
\end{figure}

To overcome this difficulty we have applied the following procedure. For all sets of simulated proton showers at fixed energy and zenith angle, we have obtained the Fisher distributions for proton showers within the region of interest delimited by the cuts on \Sem and \Smu as defined in Section\,\ref{sec:discrimination} (see also Fig.\,\ref{fig:FishCut}). The cumulative of these distributions (number of events above a Fisher value) are then obtained and normalized to one. This procedure gives the proton background selection efficiency, $\varepsilon_p$ or the proton contamination fraction as a function of the Fisher value. A few examples are shown in Fig.\,\ref{fig:four_graphs} in the Appendix. An exponential fit to the tail of the cumulative proton distributions is performed and used to extrapolate to higher background rejection factors (smaller contamination fractions $\varepsilon_p$) where the limited statistics of the proton simulations did not populate the tails of the distributions. 
The Fisher value cumulative distribution for each zenith angle is then obtained by combining the cumulative distributions for all proton energies, weighting according to their relative contribution to the cosmic-ray flux assuming a power-law $E^{-3}$ spectrum.
For each proton selection efficiency $\varepsilon_{p}$, the matching Fisher value is extracted from the cumulative of the corresponding zenith angle and taken as the Fisher cut value. 
In this way the neutrino event rate above the Fisher cut is estimated as a function of $\varepsilon_{p}$, ranging from $10^{-14}$ to $10^{-1}$, as shown in Fig.~\ref{fig:EventRatesVsEpsilonP}. The plot suggests that an electron neutrino event rate of $\sim 0.3$ per year can be achieved with proton background contamination smaller than $\sim 0.005$ per year.  The $1$-sigma uncertainty of the exponential fit can be used to evaluate the corresponding uncertainty on the number of neutrinos as a function of $\varepsilon_{p}$, shown as a band in the top panel of Fig.\,\ref{fig:EventRatesVsEpsilonP}. From this exercise, it can be seen that while the uncertainty on the number of expected neutrinos increases as $\varepsilon_{p}$ decreases, it is at maximum a factor of four for a quasi background-free ($\varepsilon_p\rightarrow 0$) experiment. In any case, for values of $\varepsilon_p$ lower than $\approx 10^{-14}$, the neutrino event rate is higher than that of background.

\section{Estimate of sensitivity for all neutrino flavours}
\label{sec:resultsall}

Until this point, this work focused exclusively on the contribution of electron neutrinos to the estimated event rate. By neglecting muon- and tau-neutrino flavors, and all anti-neutrinos, the estimate presented constitutes a lower limit to the number of neutrinos a gamma-ray ground-based array may be capable of detecting. The estimated event rate for all neutrino and anti-neutrino flavors presented here, is achieved by taking advantage of the effective mass of the array for electron-neutrinos computed in Section~\ref{sec:method} explicitly for charged (CC) and neutral-current (NC) interactions, and denoted here as $\cc$ and $\nc$ respectively. Combining these quantities with the corresponding neutrino-air interaction properties, allows us to conservatively estimate the number of expected neutrinos for all flavors and interaction channels. As we are considering astrophysical neutrinos, the expected number of electron, muon and tau neutrinos are assumed to be in the ratio 1:1:1 after oscillation over cosmological distances. Moreover, the same amount of anti-neutrinos is expected. What might change is the ability to distinguish a given species of neutrino-induced showers from the background i.e. the identification efficiency $\varepsilon$ and hence the effective mass, which can be assessed based on some qualitative arguments on the characteristics of the neutrino interaction with the Earth's atmosphere.

Firstly, the effective mass of the array when accounting only for neutral current interactions is expected to be the same for all neutrino flavors and hence equal to that of $\nu_e$ NC interactions. All neutrino flavors produce the same type of hadronic shower in a NC interaction, carrying on average the same fraction of the neutrino energy. Moreover, the only difference to the Feynman diagrams responsible for the bulk of the cross-section is the neutrino mass that can be considered negligible at the very-high energies involved. As a consequence, for all neutrino flavors the expected event rate is assumed to be proportional to $\snc\,\nc$ with $\snc$ the NC-interaction cross-section. 

For the case of the muon neutrino, the charged-current interaction will induce a hadronic shower and an energetic muon. One single muon is unlikely to be detected in a sparse array, so we will only consider the hadronic cascade. Again, given the extreme primary energies, the energy distribution of the secondaries arising from the hadronic vertex of the interaction is very similar to the one of an electron-neutrino (and an emerging fast electron) or a neutral current interaction. Hence, conservatively, we will assume that the effective mass of the array to muon neutrinos for the CC interaction is the same as the one of the electron-neutrinos for the neutral current interaction estimated before. This yields the expected number of CC-interacting muon-neutrinos proportional to $\scc\,\nc$, with $\scc$ the CC interaction cross-section.
It should be noted again that this is a conservative assumption, as the muon produced in a CC interaction could radiate an energetic photon via bremsstrahlung leading to the production of an electromagnetic shower that would increase the detection probability.

The tau-neutrino charged-current interaction produces a hadronic cascade plus a high-energy tau. In the atmosphere, the tau-lepton will travel on average between $\sim 5$ m and $\sim 5$ km at energies between 100 TeV and 100 PeV before decaying. The decay of the tau can either produce hadrons ($\sim 65\%$ of the time) and electrons ($\sim 17\%$ of the time) that will lead to \emph{young} cascades of particles. Muons can also be produced in the decay ($\sim 17\%$ of the time), that will be essentially undetectable, as discussed before. In this work, we have assumed that only the hadronic particles, directly emerging from the collision of the tau neutrino with the atmosphere, will produce a detectable shower, i.e. we neglect the decay of the $\tau$ lepton, and assume that the effective mass of the detector is the same as in the case of neutral-current interactions, with the expected number of CC-interacting tau neutrinos being proportional $\scc\,\nc$. 
We stress that this is conservative and that a more accurate calculation of the number of expected tau neutrinos would be clearly above this estimate.

The assumptions above can be applied to anti-neutrinos $\bar\nu$, given the high energy of the involved interactions. The $\bar\nu$-air interaction properties will be similar, leading to air showers with essentially the same general properties leading to similar \Sem and \Smu, the main parameters of this analysis. Additionally, above $100\,$TeV, neutrino-air and anti-neutrino-air cross-sections are very close. Nonetheless, we have used the exact values for each energy.
Consequently, the inclusion of anti-neutrinos would likely increase the expected event rate for all neutrinos by a factor 2. 

The total expected event rate would be additionally increased due to the resonant channel for the electron anti-neutrinos $\bar{\nu}_e$. Around $E_{\bar\nu} \sim 6.3\,$PeV, electron anti-neutrinos can interact with the air atomic electrons producing a real $W^-$ boson -- the so-called Glashow resonance. This resonance has in fact been observed by the IceCube neutrino observatory~\cite{IceCubeGlashow}, and represents an important contribution to the expected neutrino event rate around such energies.

In this case, the total number of expected $\bar\nu_e$-induced events, can be assumed to be proportional to $\snc\,\nc + \scc\,\cc + \sigma_G\,M_{\bar\nu_e}(\mathrm{W})$, where we denote $M_{\bar\nu_e}(\mathrm{W})$ as the effective mass for resonant anti-neutrino interactions, and $\sigma_G(E_{\bar\nu})$ is the Glashow resonance cross-section, which is a function of the anti-neutrino energy. The $W$-boson decays into hadronic particles or a lepton. Following the above considerations, $M(W)$ can be approximated as, 
\begin{eqnarray}
M_{\bar\nu_e}(\mathrm{W})\simeq\sfrac{1}{9} \cc + \sfrac{2}{3} \nc + \\ \sfrac{1}{9} \left(BR_{\tau \rightarrow e} \cc + BR_{\tau\rightarrow\rm had} \nc \right),
\label{eq:MeffGlashow}
\end{eqnarray}
where we have used the approximation that the effective mass of the array for the produced electron in the decay of the $W$ is equal $\cc$ and for hadronic final states, it follows $\nc$. The fractions accompanying the effective masses in Eq.\,(\ref{eq:MeffGlashow}) account for the (approximate) branching ratios of the $W$-boson branching ratios ($BR$) to electrons ($\sim 0.11$), hadrons ($\sim 0.68$) and $\tau$-leptons ($\sim 0.11$) with $BR_{\tau \rightarrow e}\sim 0.17$ and $ BR_{\tau\rightarrow\mathrm{had}}\sim 0.65$ denoting the tau branching ratios into electron and hadronic particles, respectively. The decay of the $W$-boson to a muon is neglected since the single muon is assumed not to produce a detectable shower, as explained before. 

\begin{figure}[ht]
	\centering
	\includegraphics[width = 0.9\linewidth]{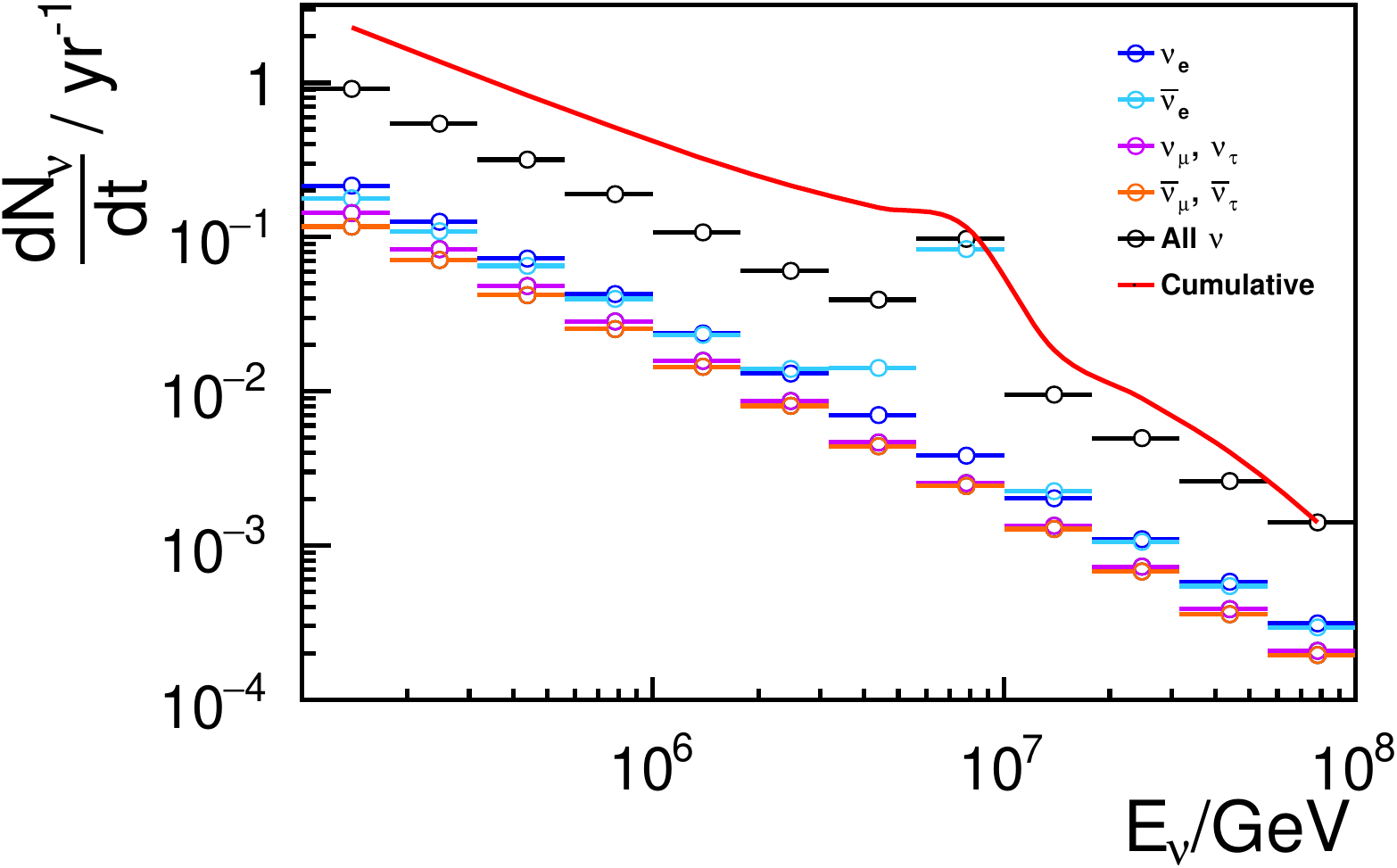}
	\caption{Event rates for all neutrino flavours within each energy bin, from $100\,{\rm TeV}$ to $100\,{\rm PeV}$. Each decade in energy is divided into 4 bins. The enhancement of the event rate at $E_\nu\sim 6.3$ PeV is due to the Glashow 
	resonant interaction of $\bar\nu_e$. The red line gives the sum of all event rates above $E_\nu$.}
	\label{fig:AllNeutrinoEventRates}
\end{figure}

With all the assumptions and approximations above, we have estimated the expected number of neutrinos per year, considering an extensive air shower array with an area of $1\,{\rm km^2}$ and a fill factor of $80\%$. This is shown in Fig.~\ref{fig:AllNeutrinoEventRates} as a function of neutrino energy and for the different neutrino flavors and channels. Accounting for the Glashow resonance of $\overline{\nu}_e$ has a noticeable impact on the total number of expected neutrinos in the energy region around $\approx 6\,$PeV. The integrated number of events per year above a given energy is also shown in Fig.\,\ref{fig:AllNeutrinoEventRates} as a red line. Integrating from $100\,$TeV up to 100 PeV, one would conservatively expect $\sim 2$ neutrino events per year. As discussed before, a more realistic array with a fill factor of $\sim 5\%$ would reduce the event rates by a factor $\lesssim 3$.
 
%-----------------------------------------------------------------------
\section{Sensitivity to upward-going Electron Neutrinos}
\label{sec:resultsup}

A study was carried out of the possibility of upward-going neutrino events contributing to the estimated event rate in a gamma-ray ground-based array of WCD. The AIRES framework was used to simulate the development of upward-going showers, 
as the version of CORSIKA code used throughout this work is unable to treat showers in dense homogeneous media such as the Earth's crust. 
We simulated upward-going showers induced by electron neutrinos, although our conclusions below apply to any type of upward-going shower. Since an electron neutrino is not a default primary particle in AIRES, we obtained the secondary products of the $\nu_e$  interaction with CORSIKA, and inject those in AIRES to obtain the longitudinal and lateral development of the shower underground.  The composition of the Earth's crust in AIRES is emulated by setting the atmosphere's composition to match that of standard soil. According to~\cite{TUEROS2010380} this medium is characterised by $\rho=1.8\,{\rm g\, cm^{-3}}$ and effective atomic number $Z=11$.
This simulation setup was utilised to inclined and very inclined up-going showers, $\theta$ ranging from $92^\circ$ to $120^\circ$ where the Earth is not opaque to neutrinos of PeV energies. We generated neutrinos with energy $E_\nu = 1\,{\rm PeV}$. The vertical height of the first interaction assumed values between $2\,{\rm m}$ and $5\, {\rm m}$ below the observation level, as showers were severely attenuated for higher depths and not sufficiently developed for smaller depths. Under each set of conditions, $1000$ showers were simulated.

\begin{figure}[ht]
         \includegraphics[width = 0.9\linewidth]{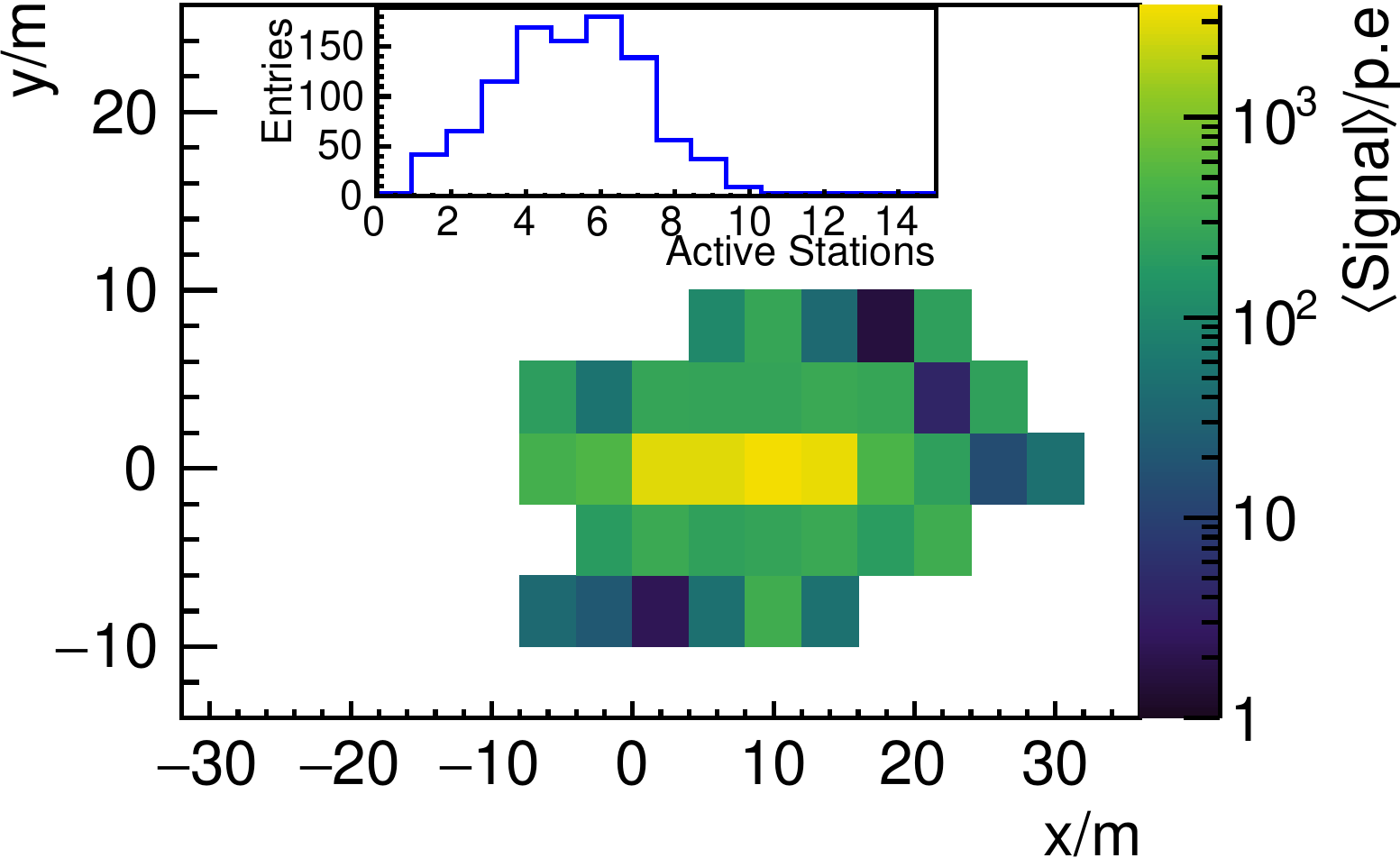}
         \caption{Average footprint produced by a shower induced by an up-going electron-neutrino with $E_\nu = 1\, {\rm PeV}$ and $\theta = 100^\circ$ interacting at a vertical depth below ground of $3\,{\rm m}$. The inset panel shows the histogram of the number of active WCD stations (stations that register signal above $10\,{\rm p.e.}$).}
     \label{fig:UpG80}
\end{figure}

The average footprint of the showers was inferred for each combination of $\theta$ and vertical depth underground. An example is presented in Fig.\,\ref{fig:UpG80} for showers with $\theta=100^\circ$ initiated at a vertical depth of $3\,{\rm m}$.
As can be seen in Fig.\,\ref{fig:UpG80}, the small dimensions of the footprints produced (of the order of a few tens of ${\rm m^2}$ in all cases), make their detection at a typical gamma-ray observatories such as LHAASO very difficult, particularly in the sparse array. The detection would eventually be possible in a compact array with larger filling factor of a gamma-ray observatory. For our nominal array with an $80\%$ filling factor, $\sim 50\%$ of the simulated events in the example shown in Fig.~\ref{fig:UpG80} have less than 5 triggered WCD stations as seen in the inset panel. Even in this case the involved effective areas would not be sufficient to perform a competitive measurement, since the shower has to be produced at less than $\sim 10$ m vertical depth below the array for it to develop before attenuating in the Earth. This limitation induces a small effective detection volume in comparison to other detection techniques such as the observation of an emerging $\tau$ decay in the atmosphere~\cite{InclinedNusPAO}. We conclude that showers induced by up-going neutrinos do not contribute significantly to the estimated event rate in the PeV energy range, explored in this work.

The Earth-skimming tau neutrino detection method consists on the observation of a shower induced by the decay of a tau lepton in the atmosphere. The tau is produced by a quasi-horizontal tau neutrino interacting in the Earth, with zenith angle between $\theta=90^\circ$ and typically $\theta\simeq 93^\circ$, corresponding to the zenith angle range where the shower can trigger an array of detectors. At the energies of interest in this work, $\sim$ PeV, the decay length of a tau lepton is of the order of 50 m, and for this reason the production of a tau-induced shower would be around 10 times more likely at PeV energies than the generation of a more upward-going shower inside the Earth that needs to initiated between 2 and 5 m depth, as explained above. However, this is partly compensated by the smaller solid angle where the shower can trigger the detector $\sim 0.22~{\rm sr}$ for $\theta\in(90^\circ,92^\circ)$ compared to $\sim 2.92~{\rm sr}$ for $\theta\in(92^\circ,120^\circ)$. On the other hand, the tau-decay induced shower produced in the atmosphere generates a footprint which will be highly dependent on the exit angle, altitude of decay and trigger conditions. One can roughly estimate a footprint of $\sim {\rm km}$ length on the array that would be more efficiently detected than the small and narrow upward-going shower produced in the larger density medium inside the Earth. As a result, the Earth-skimming technique would be, in principle, more efficient in relative terms than the detection of the upward-going showers discussed here. A more quantitative evaluation of the impact of the Earth-skimming tau neutrino channel on the total neutrino event rate requires a detailed simulation of the trigger efficiency of the EAS array to quasi-horizontal atmospheric showers, possibly considering the topography of the site, which is beyond the scope of this work. Our results in this respect should be regarded as conservative.

\section{Final remarks and Conclusions}
\label{sec:conclusions}

In this work, we have investigated the possibility of using gamma-ray wide field-of-view observatories to detect showers induced by astrophysical neutrinos in the 100 TeV to 100 PeV energy range. The discrimination from the overwhelming cosmic-ray-induced background is achieved through the detection of inclined showers and inspecting the balance between their electromagnetic and muonic content of the shower at the ground, two observables that are typically accessible in gamma-ray experiments and used for photon-hadron discrimination.
An end-to-end simulation procedure, emulating the detector response, was applied to electron neutrino events and conservatively extrapolated for the remaining neutrino and anti-neutrino species and interaction channels. The expected number of neutrinos observed through this method in an array with an effective area of $1\,{\rm km^2}$, for energies above $100\,$TeV is around 2 per year. This is not a considerable number, particularly when compared with dedicated experiments working in the same energy range, such as IceCube, which sees a few tens of events per year. Nonetheless, in the context of multi-messenger science, and the pursuit of these events, it is not negligible either. Note that gamma-ray observatories are already operating, or will be in the near future, so the potential gain of these additional events is essentially for free.
Moreover, this measurement was performed assuming a diffusive neutrino background implying that the detected neutrinos could be used to alert other experiments with a few minutes latency. 

In this work, it was also demonstrated that, while a very sensitive detection channel at very high energies, the use of upward-going events does not add much to the expected neutrino event rate due to the reduced size of the shower footprint and the relatively shallow depths of neutrino interaction needed for the shower developing underground to arrive at the array.

The number of expected neutrinos could benefit from the topography surrounding the experiments, such as mountains, as suggested in~\cite{mountains_auger, mountains_hawc}. These experiments are usually placed at high altitudes on plateaus at the foot of mountains. A shower whose reconstructed direction coincides is compatible with emerging from inside a mountain is a clean evidence of a neutrino-induced event, although the estimated rates are small.

Finally, this work aims to be a proof-of-concept, and more sophisticated analyses that could lead to higher counts are naturally envisaged. These analyses are experiment dependent, and this work shows that it is a compelling line of research to be pursued by at $\mathrm{km}^2$-scale, gamma-ray, ground-based observatories such as those pursuing PeV gamma-ray Astronomy.

\section*{Acknowledgments}
We would like to thank to Sofia Andringa and Enrique Zas for useful discussions and suggestions, and Ioana Mari\c{s} for carefully reading the manuscript.
This work has been financed by national funds through FCT - Fundação para a Ciência e a Tecnologia, I.P., under project PTDC/FIS-PAR/4300/2020. R.~C.\ is grateful for the financial support by OE - Portugal, FCT, I. P., under DL57/2016/cP1330/cT0002. This work has received financial support from Xunta de Galicia (Centro singular de investigación de Galicia accreditation 2019-2022), by European Union ERDF, and by the “María de Maeztu” Units of Excellence program MDM-2016-0692 and the Spanish Research State Agency, and from Ministerio de Ciencia e Innovaci\'on PID2019-105544GB-I00 and RED2018-102661-T (RENATA).

\bibliography{references}% Produces the bibliography via BibTeX.

\begin{thebibliography}{25}%
\makeatletter
\providecommand \@ifxundefined [1]{%
 \@ifx{#1\undefined}
}%
\providecommand \@ifnum [1]{%
 \ifnum #1\expandafter \@firstoftwo
 \else \expandafter \@secondoftwo
 \fi
}%
\providecommand \@ifx [1]{%
 \ifx #1\expandafter \@firstoftwo
 \else \expandafter \@secondoftwo
 \fi
}%
\providecommand \natexlab [1]{#1}%
\providecommand \enquote  [1]{``#1''}%
\providecommand \bibnamefont  [1]{#1}%
\providecommand \bibfnamefont [1]{#1}%
\providecommand \citenamefont [1]{#1}%
\providecommand \href@noop [0]{\@secondoftwo}%
\providecommand \href [0]{\begingroup \@sanitize@url \@href}%
\providecommand \@href[1]{\@@startlink{#1}\@@href}%
\providecommand \@@href[1]{\endgroup#1\@@endlink}%
\providecommand \@sanitize@url [0]{\catcode `\\12\catcode `\$12\catcode
  `\&12\catcode `\#12\catcode `\^12\catcode `\_12\catcode `\%12\relax}%
\providecommand \@@startlink[1]{}%
\providecommand \@@endlink[0]{}%
\providecommand \url  [0]{\begingroup\@sanitize@url \@url }%
\providecommand \@url [1]{\endgroup\@href {#1}{\urlprefix }}%
\providecommand \urlprefix  [0]{URL }%
\providecommand \Eprint [0]{\href }%
\providecommand \doibase [0]{http://dx.doi.org/}%
\providecommand \selectlanguage [0]{\@gobble}%
\providecommand \bibinfo  [0]{\@secondoftwo}%
\providecommand \bibfield  [0]{\@secondoftwo}%
\providecommand \translation [1]{[#1]}%
\providecommand \BibitemOpen [0]{}%
\providecommand \bibitemStop [0]{}%
\providecommand \bibitemNoStop [0]{.\EOS\space}%
\providecommand \EOS [0]{\spacefactor3000\relax}%
\providecommand \BibitemShut  [1]{\csname bibitem#1\endcsname}%
\let\auto@bib@innerbib\@empty
%</preamble>
\bibitem [{\citenamefont {Vargas}(2020)}]{HAWC}%
  \BibitemOpen
  \bibfield  {author} {\bibinfo {author} {\bibfnamefont {H.~L.}\ \bibnamefont
  {Vargas}} (\bibinfo {collaboration} {HAWC}),\ }\href {\doibase
  10.22323/1.358.0940} {\bibfield  {journal} {\bibinfo  {journal} {PoS}\
  }\textbf {\bibinfo {volume} {ICRC2019}},\ \bibinfo {pages} {940} (\bibinfo
  {year} {2020})}\BibitemShut {NoStop}%
\bibitem [{\citenamefont {Bai~et al.}(2019)}]{LHAASOLayout}%
  \BibitemOpen
  \bibfield  {author} {\bibinfo {author} {\bibfnamefont {X.}~\bibnamefont
  {Bai~et al.}},\ }\href@noop {} {\enquote {\bibinfo {title} {{The Large High
  Altitude Air Shower Observatory (LHAASO) Science White Paper}},}\ } (\bibinfo
  {year} {2019}),\ \Eprint {http://arxiv.org/abs/1905.02773} {arXiv:1905.02773
  [astro-ph.HE]} \BibitemShut {NoStop}%
\bibitem [{\citenamefont {Abreu}\ \emph {et~al.}(2019)\citenamefont {Abreu}
  \emph {et~al.}}]{SWGOFuture}%
  \BibitemOpen
  \bibfield  {author} {\bibinfo {author} {\bibfnamefont {P.}~\bibnamefont
  {Abreu}} \emph {et~al.},\ }\href@noop {} {\  (\bibinfo {year} {2019})},\
  \Eprint {http://arxiv.org/abs/1907.07737} {arXiv:1907.07737 [astro-ph.IM]}
  \BibitemShut {NoStop}%
\bibitem [{\citenamefont {Cao}\ \emph {et~al.}(2021)\citenamefont {Cao} \emph
  {et~al.}}]{LHAASOPeV}%
  \BibitemOpen
  \bibfield  {author} {\bibinfo {author} {\bibfnamefont {Z.}~\bibnamefont
  {Cao}} \emph {et~al.},\ }\href {\doibase 10.1038/s41586-021-03498-z}
  {\bibfield  {journal} {\bibinfo  {journal} {Nature}\ }\textbf {\bibinfo
  {volume} {594}},\ \bibinfo {pages} {33} (\bibinfo {year} {2021})}\BibitemShut
  {NoStop}%
\bibitem [{\citenamefont {Aartsen}\ \emph {et~al.}(2020)\citenamefont {Aartsen}
  \emph {et~al.}}]{IceCubeFlux}%
  \BibitemOpen
  \bibfield  {author} {\bibinfo {author} {\bibfnamefont {M.~G.}\ \bibnamefont
  {Aartsen}} \emph {et~al.} (\bibinfo {collaboration} {IceCube}),\ }\href
  {\doibase 10.1103/PhysRevLett.125.121104} {\bibfield  {journal} {\bibinfo
  {journal} {Phys. Rev. Lett.}\ }\textbf {\bibinfo {volume} {125}},\ \bibinfo
  {pages} {121104} (\bibinfo {year} {2020})},\ \Eprint
  {http://arxiv.org/abs/2001.09520} {arXiv:2001.09520 [astro-ph.HE]}
  \BibitemShut {NoStop}%
\bibitem [{\citenamefont {Aartsen}\ \emph
  {et~al.}(2021{\natexlab{a}})\citenamefont {Aartsen} \emph
  {et~al.}}]{IceCube-Gen2}%
  \BibitemOpen
  \bibfield  {author} {\bibinfo {author} {\bibfnamefont {M.~G.}\ \bibnamefont
  {Aartsen}} \emph {et~al.} (\bibinfo {collaboration} {IceCube-Gen2}),\ }\href
  {\doibase 10.1088/1361-6471/abbd48} {\bibfield  {journal} {\bibinfo
  {journal} {J. Phys. G}\ }\textbf {\bibinfo {volume} {48}},\ \bibinfo {pages}
  {060501} (\bibinfo {year} {2021}{\natexlab{a}})},\ \Eprint
  {http://arxiv.org/abs/2008.04323} {arXiv:2008.04323 [astro-ph.HE]}
  \BibitemShut {NoStop}%
\bibitem [{\citenamefont {Abdalla}\ \emph {et~al.}(2021)\citenamefont {Abdalla}
  \emph {et~al.}}]{CTAFundPhys}%
  \BibitemOpen
  \bibfield  {author} {\bibinfo {author} {\bibfnamefont {H.}~\bibnamefont
  {Abdalla}} \emph {et~al.} (\bibinfo {collaboration} {CTA}),\ }\href {\doibase
  10.1088/1475-7516/2021/02/048} {\bibfield  {journal} {\bibinfo  {journal}
  {JCAP}\ }\textbf {\bibinfo {volume} {02}},\ \bibinfo {pages} {048} (\bibinfo
  {year} {2021})},\ \Eprint {http://arxiv.org/abs/2010.01349} {arXiv:2010.01349
  [astro-ph.HE]} \BibitemShut {NoStop}%
\bibitem [{\citenamefont {Liu}\ \emph {et~al.}(2022)\citenamefont {Liu} \emph
  {et~al.}}]{FermiGW}%
  \BibitemOpen
  \bibfield  {author} {\bibinfo {author} {\bibfnamefont {Y.}~\bibnamefont
  {Liu}} \emph {et~al.} (\bibinfo {collaboration} {Fermi-LAT}),\ }\href
  {\doibase 10.1126/science.abm3231} {\bibfield  {journal} {\bibinfo  {journal}
  {Science}\ }\textbf {\bibinfo {volume} {376}},\ \bibinfo {pages} {abm3231}
  (\bibinfo {year} {2022})},\ \Eprint {http://arxiv.org/abs/2204.05226}
  {arXiv:2204.05226 [astro-ph.HE]} \BibitemShut {NoStop}%
\bibitem [{Note1()}]{Note1}%
  \BibitemOpen
  \bibinfo {note} {In this context, the fill factor is the total detector
  sensitive area over the shower sampling area (size of the
  array).}\BibitemShut {Stop}%
\bibitem [{\citenamefont {Abreu}\ \emph {et~al.}(2011)\citenamefont {Abreu}
  \emph {et~al.}}]{PierreAuger:2011cpc}%
  \BibitemOpen
  \bibfield  {author} {\bibinfo {author} {\bibfnamefont {P.}~\bibnamefont
  {Abreu}} \emph {et~al.} (\bibinfo {collaboration} {Pierre Auger}),\ }\href
  {\doibase 10.1103/PhysRevD.84.122005} {\bibfield  {journal} {\bibinfo
  {journal} {Phys. Rev. D}\ }\textbf {\bibinfo {volume} {84}},\ \bibinfo
  {pages} {122005} (\bibinfo {year} {2011})},\ \bibinfo {note} {[Erratum:
  Phys.Rev.D 84, 029902 (2011)]},\ \Eprint {http://arxiv.org/abs/1202.1493}
  {arXiv:1202.1493 [astro-ph.HE]} \BibitemShut {NoStop}%
\bibitem [{\citenamefont {Aab~et al.}(2020)}]{PAOUHENus}%
  \BibitemOpen
  \bibfield  {author} {\bibinfo {author} {\bibfnamefont {A.}~\bibnamefont
  {Aab~et al.}},\ }\href {\doibase 10.3847/1538-4357/abb476} {\bibfield
  {journal} {\bibinfo  {journal} {The Astrophysical Journal}\ }\textbf
  {\bibinfo {volume} {902}},\ \bibinfo {pages} {105} (\bibinfo {year}
  {2020})}\BibitemShut {NoStop}%
\bibitem [{\citenamefont {Aramo~et al.}(2005)}]{InclinedNusPAO}%
  \BibitemOpen
  \bibfield  {author} {\bibinfo {author} {\bibfnamefont {C.}~\bibnamefont
  {Aramo~et al.}},\ }\href {\doibase 10.1016/j.astropartphys.2004.11.008}
  {\bibfield  {journal} {\bibinfo  {journal} {Astroparticle Physics}\ }\textbf
  {\bibinfo {volume} {23}},\ \bibinfo {pages} {65–77} (\bibinfo {year}
  {2005})}\BibitemShut {NoStop}%
\bibitem [{\citenamefont {Heck}\ \emph {et~al.}(1998)\citenamefont {Heck},
  \citenamefont {Capdevielle}, \citenamefont {Schatz}, \citenamefont {Thouw},\
  and\ \citenamefont {Gmbh}}]{CORSIKA}%
  \BibitemOpen
  \bibfield  {author} {\bibinfo {author} {\bibfnamefont {D.}~\bibnamefont
  {Heck}}, \bibinfo {author} {\bibfnamefont {J.~N.}\ \bibnamefont
  {Capdevielle}}, \bibinfo {author} {\bibfnamefont {G.}~\bibnamefont {Schatz}},
  \bibinfo {author} {\bibfnamefont {T.}~\bibnamefont {Thouw}}, \ and\ \bibinfo
  {author} {\bibfnamefont {F.~K.}\ \bibnamefont {Gmbh}},\ }\href@noop {}
  {\enquote {\bibinfo {title} {{CORSIKA: A Monte Carlo Code to Simulate
  Extensive Air Showers, Report FZKA 6019, Forschungszentrum Karlsruhe}},}\ }
  (\bibinfo {year} {1998})\BibitemShut {NoStop}%
\bibitem [{\citenamefont {Sciutto}(1999)}]{AIRES}%
  \BibitemOpen
  \bibfield  {author} {\bibinfo {author} {\bibfnamefont {S.~J.}\ \bibnamefont
  {Sciutto}},\ }\href {\doibase 10.13140/RG.2.2.12566.40002} {\  (\bibinfo
  {year} {1999}),\ 10.13140/RG.2.2.12566.40002},\ \Eprint
  {http://arxiv.org/abs/astro-ph/9911331} {arXiv:astro-ph/9911331} \BibitemShut
  {NoStop}%
\bibitem [{\citenamefont {Concei\c{c}\~ao}\ \emph {et~al.}(2021)\citenamefont
  {Concei\c{c}\~ao}, \citenamefont {Gonz\'alez}, \citenamefont {Guill\'en},
  \citenamefont {Pimenta},\ and\ \citenamefont {Tom\'e}}]{wcd4pmt}%
  \BibitemOpen
  \bibfield  {author} {\bibinfo {author} {\bibfnamefont {R.}~\bibnamefont
  {Concei\c{c}\~ao}}, \bibinfo {author} {\bibfnamefont {B.~S.}\ \bibnamefont
  {Gonz\'alez}}, \bibinfo {author} {\bibfnamefont {A.}~\bibnamefont
  {Guill\'en}}, \bibinfo {author} {\bibfnamefont {M.}~\bibnamefont {Pimenta}},
  \ and\ \bibinfo {author} {\bibfnamefont {B.}~\bibnamefont {Tom\'e}},\ }\href
  {\doibase 10.1140/epjc/s10052-021-09312-4} {\bibfield  {journal} {\bibinfo
  {journal} {Eur. Phys. J. C}\ }\textbf {\bibinfo {volume} {81}},\ \bibinfo
  {pages} {542} (\bibinfo {year} {2021})},\ \Eprint
  {http://arxiv.org/abs/2101.10109} {arXiv:2101.10109 [physics.ins-det]}
  \BibitemShut {NoStop}%
\bibitem [{\citenamefont {Agostinelli~et al.}(2003)}]{Geant4}%
  \BibitemOpen
  \bibfield  {author} {\bibinfo {author} {\bibfnamefont {S.}~\bibnamefont
  {Agostinelli~et al.}},\ }\href {\doibase 10.1016/S0168-9002(03)01368-8}
  {\bibfield  {journal} {\bibinfo  {journal} {Nucl. Instrum. Meth.}\ }\textbf
  {\bibinfo {volume} {A506}},\ \bibinfo {pages} {250} (\bibinfo {year}
  {2003})}\BibitemShut {NoStop}%
\bibitem [{\citenamefont {Hocker}\ \emph {et~al.}(2007)\citenamefont {Hocker}
  \emph {et~al.}}]{TMVA}%
  \BibitemOpen
  \bibfield  {author} {\bibinfo {author} {\bibfnamefont {A.}~\bibnamefont
  {Hocker}} \emph {et~al.},\ }\href@noop {} {\  (\bibinfo {year} {2007})},\
  \Eprint {http://arxiv.org/abs/physics/0703039} {arXiv:physics/0703039}
  \BibitemShut {NoStop}%
\bibitem [{\citenamefont {{Connolly}}\ \emph {et~al.}(2011)\citenamefont
  {{Connolly}}, \citenamefont {{Thorne}},\ and\ \citenamefont
  {{Waters}}}]{NuXSecs}%
  \BibitemOpen
  \bibfield  {author} {\bibinfo {author} {\bibfnamefont {A.}~\bibnamefont
  {{Connolly}}}, \bibinfo {author} {\bibfnamefont {R.~S.}\ \bibnamefont
  {{Thorne}}}, \ and\ \bibinfo {author} {\bibfnamefont {D.}~\bibnamefont
  {{Waters}}},\ }\href {\doibase 10.1103/PhysRevD.83.113009} {\bibfield
  {journal} {\bibinfo  {journal} {Physical Review D}\ }\textbf {\bibinfo
  {volume} {83}},\ \bibinfo {eid} {113009} (\bibinfo {year}
  {2011})}\BibitemShut {NoStop}%
\bibitem [{Note2()}]{Note2}%
  \BibitemOpen
  \bibinfo {note} {$E_\nu $ and $\theta $ are fixed for each case.}\BibitemShut
  {Stop}%
\bibitem [{\citenamefont {Aab}\ \emph {et~al.}(2020)\citenamefont {Aab} \emph
  {et~al.}}]{conic_geom}%
  \BibitemOpen
  \bibfield  {author} {\bibinfo {author} {\bibfnamefont {A.}~\bibnamefont
  {Aab}} \emph {et~al.} (\bibinfo {collaboration} {Pierre Auger}),\ }\href
  {\doibase 10.1088/1748-0221/15/10/P10021} {\bibfield  {journal} {\bibinfo
  {journal} {JINST}\ }\textbf {\bibinfo {volume} {15}},\ \bibinfo {pages}
  {P10021} (\bibinfo {year} {2020})},\ \Eprint
  {http://arxiv.org/abs/2007.09035} {arXiv:2007.09035 [astro-ph.IM]}
  \BibitemShut {NoStop}%
\bibitem [{\citenamefont {Aab}\ \emph {et~al.}(2015)\citenamefont {Aab} \emph
  {et~al.}}]{AugerInclined}%
  \BibitemOpen
  \bibfield  {author} {\bibinfo {author} {\bibfnamefont {A.}~\bibnamefont
  {Aab}} \emph {et~al.} (\bibinfo {collaboration} {Pierre Auger}),\ }\href
  {\doibase 10.1103/PhysRevD.91.059901, 10.1103/PhysRevD.91.032003} {\bibfield
  {journal} {\bibinfo  {journal} {Phys. Rev.}\ }\textbf {\bibinfo {volume}
  {D91}},\ \bibinfo {pages} {032003} (\bibinfo {year} {2015})},\ \bibinfo
  {note} {[Erratum: Phys. Rev.D91,no.5,059901(2015)]},\ \Eprint
  {http://arxiv.org/abs/1408.1421} {arXiv:1408.1421 [astro-ph.HE]} \BibitemShut
  {NoStop}%
%%CITATION = ARXIV:1408.1421;%%
\bibitem [{\citenamefont {Aartsen}\ \emph
  {et~al.}(2021{\natexlab{b}})\citenamefont {Aartsen} \emph
  {et~al.}}]{IceCubeGlashow}%
  \BibitemOpen
  \bibfield  {author} {\bibinfo {author} {\bibfnamefont {M.~G.}\ \bibnamefont
  {Aartsen}} \emph {et~al.} (\bibinfo {collaboration} {IceCube}),\ }\href
  {\doibase 10.1038/s41586-021-03256-1} {\bibfield  {journal} {\bibinfo
  {journal} {Nature}\ }\textbf {\bibinfo {volume} {591}},\ \bibinfo {pages}
  {220} (\bibinfo {year} {2021}{\natexlab{b}})},\ \bibinfo {note} {[Erratum:
  Nature 592, E11 (2021)]},\ \Eprint {http://arxiv.org/abs/2110.15051}
  {arXiv:2110.15051 [hep-ex]} \BibitemShut {NoStop}%
\bibitem [{\citenamefont {Tueros}\ and\ \citenamefont
  {Sciutto}(2010)}]{TUEROS2010380}%
  \BibitemOpen
  \bibfield  {author} {\bibinfo {author} {\bibfnamefont {M.}~\bibnamefont
  {Tueros}}\ and\ \bibinfo {author} {\bibfnamefont {S.}~\bibnamefont
  {Sciutto}},\ }\href {\doibase https://doi.org/10.1016/j.cpc.2009.09.022}
  {\bibfield  {journal} {\bibinfo  {journal} {Computer Physics Communications}\
  }\textbf {\bibinfo {volume} {181}},\ \bibinfo {pages} {380} (\bibinfo {year}
  {2010})}\BibitemShut {NoStop}%
\bibitem [{\citenamefont {Abreu}\ \emph {et~al.}(2013)\citenamefont {Abreu}
  \emph {et~al.}}]{mountains_auger}%
  \BibitemOpen
  \bibfield  {author} {\bibinfo {author} {\bibfnamefont {P.}~\bibnamefont
  {Abreu}} \emph {et~al.} (\bibinfo {collaboration} {Pierre Auger}),\ }\href
  {\doibase 10.1155/2013/708680} {\bibfield  {journal} {\bibinfo  {journal}
  {Adv. High Energy Phys.}\ }\textbf {\bibinfo {volume} {2013}},\ \bibinfo
  {pages} {708680} (\bibinfo {year} {2013})},\ \Eprint
  {http://arxiv.org/abs/1304.1630} {arXiv:1304.1630 [astro-ph.HE]} \BibitemShut
  {NoStop}%
\bibitem [{\citenamefont {Albert}\ \emph {et~al.}(2022)\citenamefont {Albert}
  \emph {et~al.}}]{mountains_hawc}%
  \BibitemOpen
  \bibfield  {author} {\bibinfo {author} {\bibfnamefont {A.}~\bibnamefont
  {Albert}} \emph {et~al.} (\bibinfo {collaboration} {HAWC}),\ }\href {\doibase
  10.1016/j.astropartphys.2021.102670} {\bibfield  {journal} {\bibinfo
  {journal} {Astropart. Phys.}\ }\textbf {\bibinfo {volume} {137}},\ \bibinfo
  {pages} {102670} (\bibinfo {year} {2022})},\ \Eprint
  {http://arxiv.org/abs/2108.07767} {arXiv:2108.07767 [hep-ex]} \BibitemShut
  {NoStop}%
\end{thebibliography}%

\appendix

\section{Fits to the Proton Fisher cumulative distribution tail}

\begin{figure*}[htp]
\subfloat[\label{subfig:a}]{%
  \includegraphics[width=0.45\textwidth]{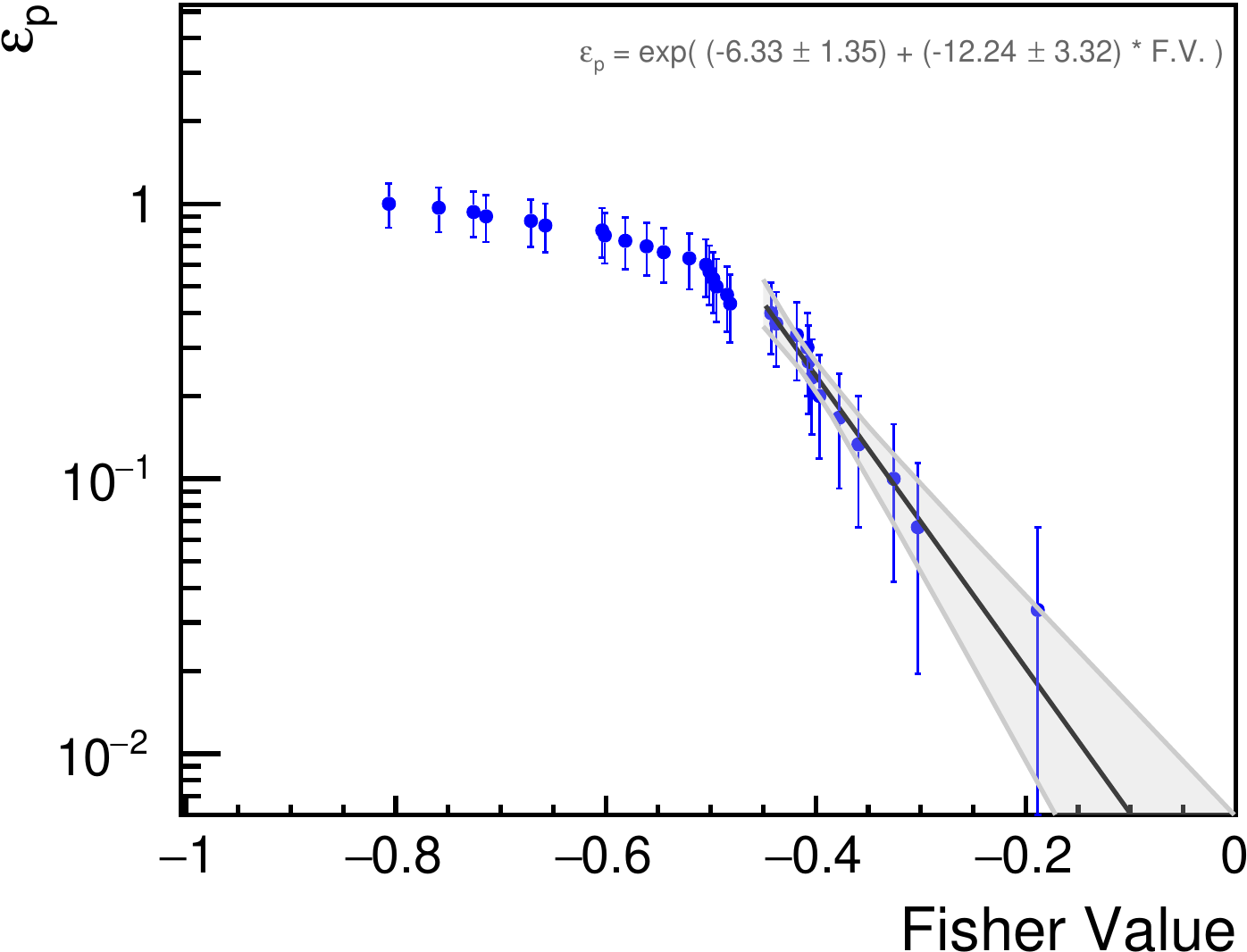}%
}
\subfloat[\label{subfig:b}]{%
  \includegraphics[width=0.45\textwidth]{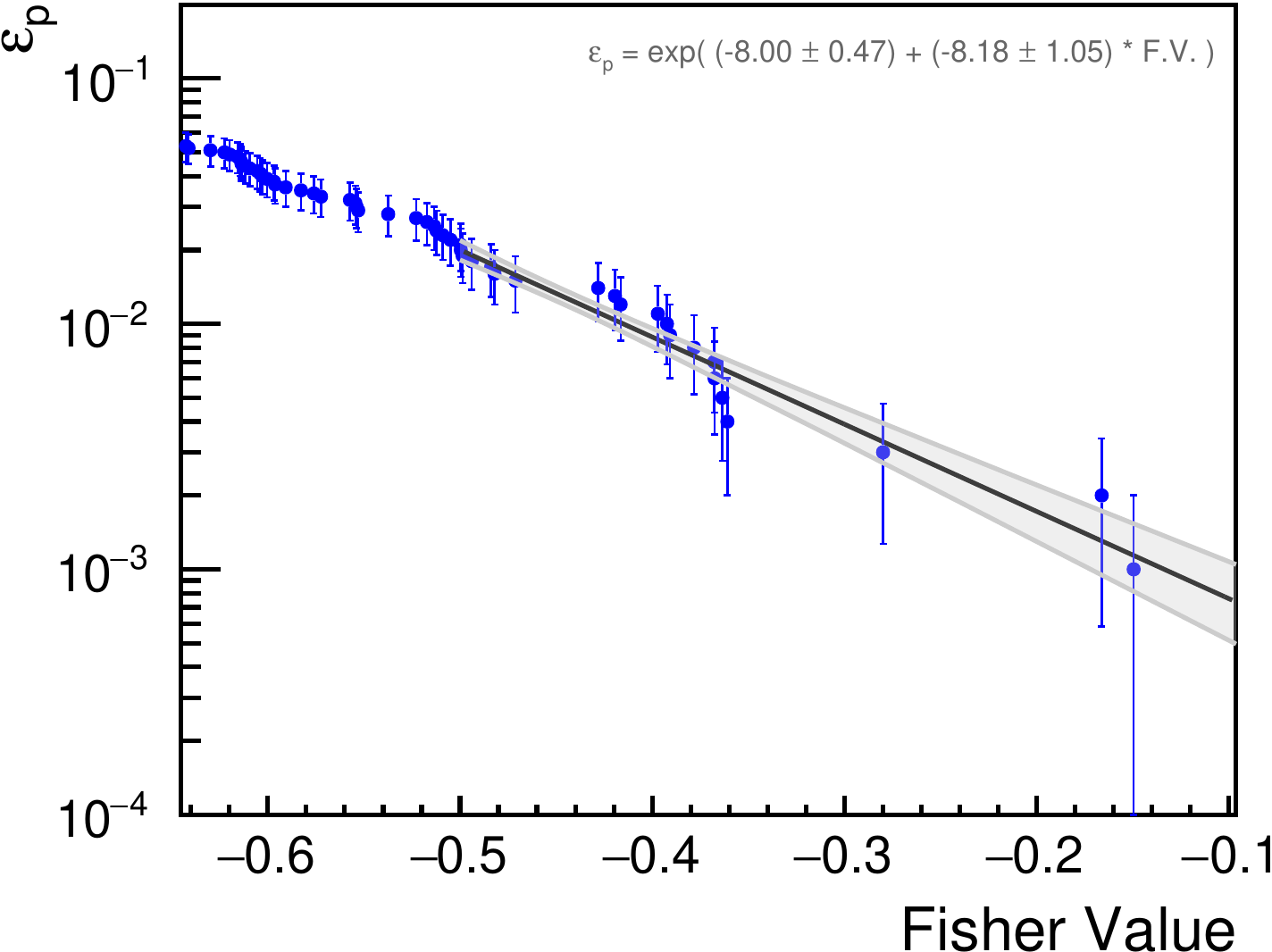}%
}\hfill
\subfloat[\label{subfig:c}]{%
  \includegraphics[width=0.45\textwidth]{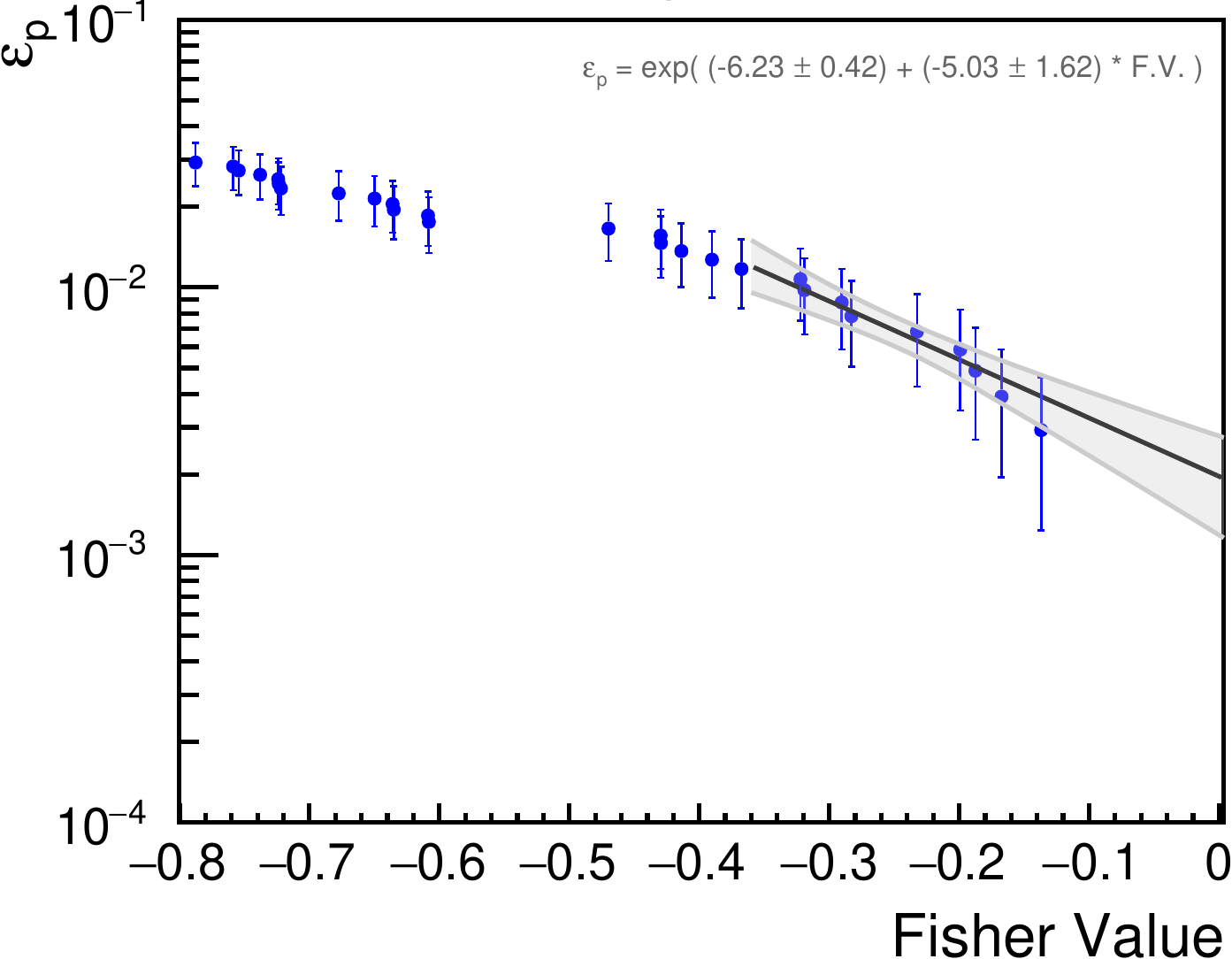}%
}
\subfloat[\label{subfig:d}]{%
  \includegraphics[width=0.45\textwidth]{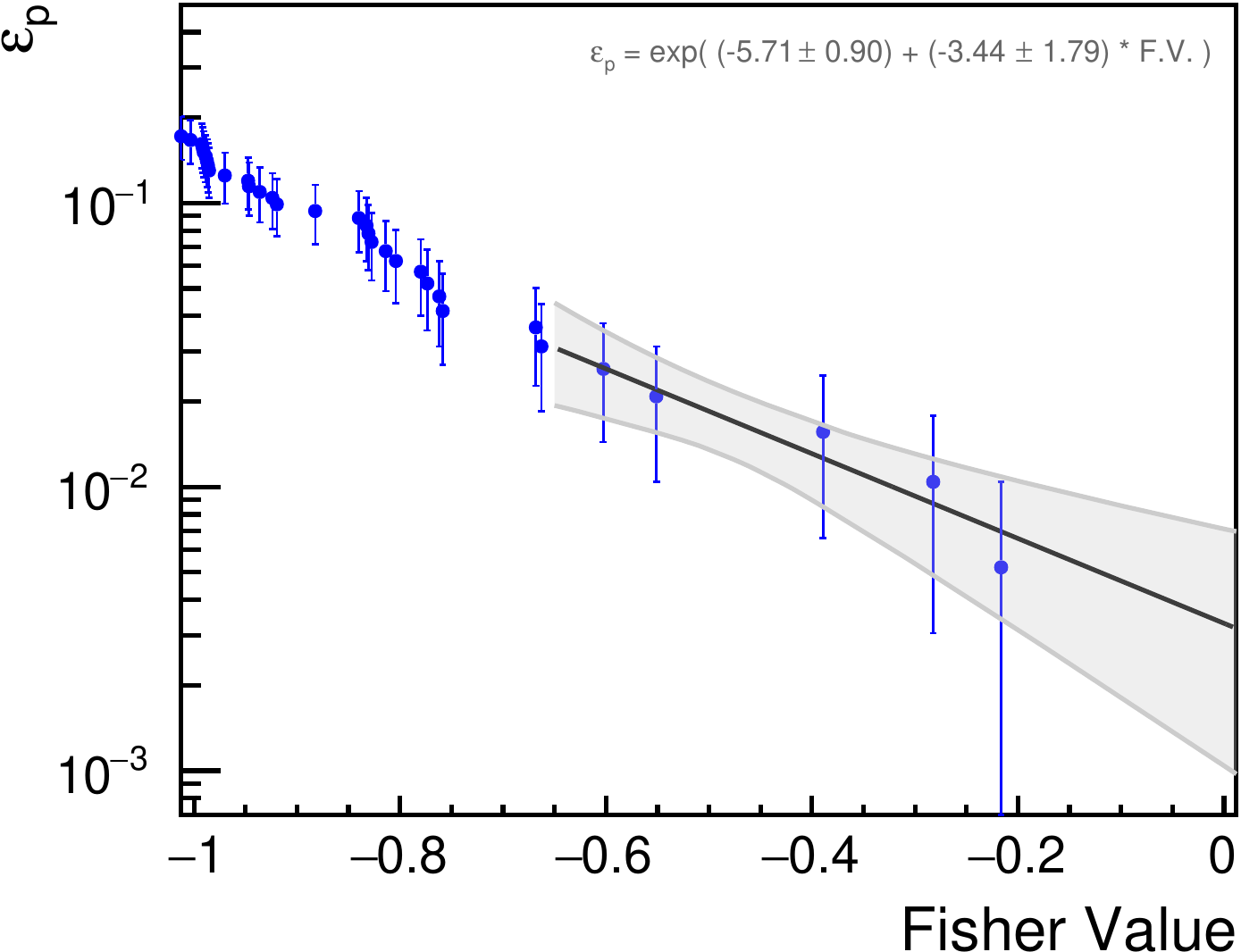}%
}\hfill
\caption{Example of the exponential fits to the tail of the cumulative proton distributions (solid black line) used to extrapolate to higher background rejection factors (smaller contamination fractions $\varepsilon_p$). The 1 sigma uncertainty of the fit corresponds to the shaded area. \textbf{(a)} $\theta = 60^{\circ},\, E_p = 10^4\,{\rm GeV}$ \textbf{(b)} $\theta = 60^{\circ},\, E_p = 10^6\,{\rm GeV}$ \textbf{(c)} $\theta = 70^{\circ},\, E_p = 10^7\,{\rm GeV}$ \textbf{(d)} $\theta = 75^{\circ},\, E_p = 10^9\,{\rm GeV}$.}\label{fig:four_graphs}
\end{figure*}

A few examples of the normalized cumulatives of the Fisher value distributions for proton showers within the region of interest are presented here. The formula of the exponential fit performed to the tail of the cumulative is presented in each figure. The exponential fit is presented as a solid black line, and the shaded area represents its 1-sigma uncertainty. This fit was used to extrapolate to higher background rejection factors (smaller contamination fractions $\varepsilon_p$).

\end{document}